\documentclass[a4paper,12pt]{article}

\usepackage{epsfig}
\usepackage{color}
\usepackage{array}

\usepackage{amsmath}
\usepackage{pstricks}
\usepackage{bm}
\usepackage{subfig}

\usepackage{graphicx}

\newcommand{\be}{\begin{equation}}
\newcommand{\ee}{\end{equation}}

\def\mc{m_\chi^2}
\def\mo{m_\omega^2}
\def\sw{\sin^2\theta_W}

\newcommand{\beq}[1] {\begin{equation}\label{#1} }
\newcommand{\eeq} {\end{equation} }
\newcommand{\bea}[1]{\begin{eqnarray}\label{#1} }
\newcommand{\eea}{\end{eqnarray}}

\def\beqn{\begin{eqnarray}}
\def\eeqn{\end{eqnarray}}

\def\beq{\begin{equation}}
\def\eeq{\end{equation}}
\def\bea{\begin{equation}}
\def\eea{\end{equation}}

\def\ga{\gamma}

\def\om{\omega}

\def\hs{\hspace}

\def\ov{\overline}

\def\ol{\overline}

\def\yyd{\left(YY^\dagger\right)_{11}}
\def\ffd{\left(FF^\dagger\right)_{11}}

\begin{document}

\begin{flushright}
OSU-HEP-10-05\\
June 5, 2010 \\
\end{flushright}

\renewcommand{\thefootnote}{\fnsymbol{footnote}}

\begin{center}
{\Large\bf Two--Loop Neutrino Mass Generation \\[0.1in]
through Leptoquarks}\\
\end{center}

\vspace{0.5cm}
\begin{center}
{\large \bf {}~K.S. Babu$^{\hs{0.5mm}}$\footnote{Email:
babu@okstate.edu} and J. Julio$^{\hs{0.5mm}}$\footnote{julio.julio@okstate.edu} }
\vspace{0.5cm}

$^a${\em Department of Physics \\ Oklahoma State University \\
Stillwater, OK 74078, USA }

\end{center}

\begin{abstract}
We present a new model of radiative neutrino mass generation wherein
TeV scale leptoguark scalars induce tiny neutrino masses as
two--loop radiative corrections.  The neutrino oscillation parameter
$\sin^2\theta_{13}$ is predicted to be close to the current experimental
limit within the model.  Rare lepton flavor violating processes
mediated by leptoquarks have an interesting pattern: $\mu
\rightarrow e \gamma$ may be suppressed, while $\mu \rightarrow 3 e$
and $\mu-e$ conversion in nuclei are within reach of the next
generation experiments.  New CP violating contributions to $B_s-
\ol{B}_s$ mixing via leptoquark box diagrams are in a range that can
explain the recently reported discrepancy with the standard model.
$D_s^- \rightarrow \ell^- \nu$ decays mediated by leptoquarks brings
theory and experiment closer, removing an observed $2\sigma$
anomaly. Muon $g-2$ receives new positive contributions, which can
resolve the discrepancy between theory and experiment.  The
leptoquarks of the model are accessible to the LHC, and their decay
branching ratios probe neutrino oscillation parameters.
\end{abstract}

\newpage

\section{Introduction}\label{sec:intro}
\renewcommand{\thefootnote}{\arabic{footnote}}
\setcounter{footnote}{0}

It has now been firmly established that neutrinos have tiny masses
and that oscillation occurs between different flavors.  The standard
paradigm that explains the small masses is the seesaw mechanism
\cite{seesaw}, which generates an effective dimension--5 operator
${\cal O}_1 = (L L HH)/M$, suppressed by the mass scale $M$ of the
heavy right--handed neutrino.  Oscillation data suggests that in
this scenario $M \sim 10^{14}$, which is well beyond the reach of
foreseeable experiments for direct scrutiny. One has to rely on
other indirect hints, such as charged lepton flavor violation in a
supersymmetric context, in order to falsify this theory.

An interesting alternative to the high scale seesaw mechanism is
to induce small neutrino masses at the loop level.  The smallness
of neutrino masses can be understood as originating from
loop and chirality suppression factors.  The simplest among this class
of models is the Zee model \cite{smirnov} where neutrino masses are
induced as one--loop radiative corrections arising from the exchange
of charged scalar bosons.  The effective operator in this model is
$LLLe^cH/M$.  To convert this to neutrino mass, a loop diagram is necessary.\footnote{This
model is now excluded by neutrino oscillation
data (see Ref. \cite{he}).} In a second class of models, neutrino
masses arise as two--loop radiative corrections via the exchange of
singly and doubly charged scalars \cite{zee1,babu}.  The effective operator of these
models is $LLL e^c L e^c/M^2$.  This model is compatible with neutrino oscillation
data.  Phenomenology of these models has
been studied in Ref. \cite{babu1}.

A classification of low dimensional effective $\Delta L = 2$ lepton
number violating operators that can lead to neutrino masses has been
given in Ref. \cite{babu-leung}.  Among these is an operator labeled
${\cal O}_8$:
\begin{equation}
{\cal O}_8 = L_\alpha \overline{e^c} ~\overline{u^c} d^c H_\beta
\epsilon^{\alpha \beta}~
\end{equation}
where $\alpha,\beta$ are $SU(2)_L$ indices, with the family
indices suppressed.   It is the purpose of this paper
to develop a renormalizable model that generates ${\cal O}_8$.  We
will show that such a model can be consistently constructed, with a variety of
testable predictions.

Operator ${\cal O}_8$ is most directly induced by the exchange of scalar
leptoquarks.\footnote{Neutrino mass generation at the one-loop level by
leptoquark exchange has been studied in Ref. \cite{hirsch}.  Neutrino
mass generation at the two--loop level in the supersymmetric standard model with specific
$R$--parity violating couplings has been studied in
Ref. \cite{tl-susy}.}  For neutrino mass generation the two quark fields and
the charged lepton field in  ${\cal O}_8$ will have to be removed,
which implies that the masses arise at the two--loop level.  The
order of magnitude of the induced neutrino masses is
\begin{equation}
m_\nu \sim \frac{m_t m_b m_\tau \mu v}{(16 \pi^2)^2 M_{\rm LQ}^4},
\end{equation}
where $\mu$ is a dimensionful coefficient of a cubic scalar
coupling, and $v= 174$ GeV
 is the electroweak vacuum expectation value (VEV). In order to generate $m_\nu \sim 0.05$ eV, it
 is clear that $M_{\rm LQ}$ must be of order TeV, which would be within reach of the LHC.

TeV scale leptoquarks can mediate a variety of flavor changing
processes.  We have examined these constraints and found the
neutrino mass generation to be self-consistent.  Our findings
include several interesting features:  (i) $\mu \rightarrow e
\gamma$ may be suppressed, because of a GIM--like cancelation, but
$\mu \rightarrow 3e$ and $\mu-e$ conversion nuclei are within reach
of the next generation of experiments.  (ii) There is a new CP
violating contribution mediated by leptoquark box diagrams in
$B_s-\ol{B}_s$ mixing, which can nicely fit the recently reported
dimuon anomaly by the D{\O} collaboration.  (iii) The 2$\sigma$ discrepancy between theory and
experiment in the leptonic decay of $D_s^\pm$ mesons can be
explained in the model, owing to new contributions from the
leptoquarks. (iv) Muon $g-2$ receives new positive contributions
which can resolve the theoretical anomaly there. (v) Neutrino
oscillation parameters can be probed in the branching ratios of the
leptoquarks. (vi) The leptoquark masses are constrained to be less
than a few TeV, which should make them accessible to the LHC.

This paper is organized as follows. In Section 2 we discuss the
model leading to two--loop neutrino mass generation and obtain the
constraints placed on the oscillation parameters. The constraints
from rare processes such as $\mu^- \to e^- \gamma$ and $\mu^- \to
e^+e^-e^-$ are presented in Section 3. In Sec. 4 we  discuss
the collider signals of leptoquarks, and in Sec. 5 we conclude.

\section{Model of two--loop neutrino mass generation}
In this section we present our model of two--loop neutrino mass generation,
and derive restrictions on the model parameters from neutrino oscillation data.
Constraints from rare processes, discussed in Sec. 3, will be used to demonstrate
the viability of the model and its predictions for neutrino oscillations.

\subsection{Model}

The gauge symmetry of our model is the same as the standard model (SM),
$SU(3)_c\times SU(2)_L\times U(1)_Y$.
In addition to the SM Higgs doublet $H(1,2,1/2)$, the scalar sector consists of the following leptoquark
multiplets:
\begin{equation}
\Omega \equiv \left(\begin{array}{c}\omega^{2/3}
\\ \omega^{-1/3}\end{array}\right) \sim (3,2,1/6),
~~\qquad\chi^{-1/3}\sim (3,1,-1/3).
\end{equation}
In general, addition of leptoquarks into the theory can cause baryon number ($B$)
violating interactions, we forbid them by assuming that
$B$ is globally conserved.
The leptoquarks have the following Yukawa interactions:
\begin{equation}
{\cal L}_{\rm Yukawa} = Y_{ij} L_i^\alpha d^c_j \Omega^\beta
\epsilon_{\alpha \beta} + F_{ij} e_i^c u_j^c \chi^{-1/3} + {\rm
h.c.} \label{interaction}
\end{equation}
Here $i,~j=1-3$ are family indices and $\alpha,~\beta$ are $SU(2)_L$
indices.  Note that these Yukawa couplings conserve both baryon
number and lepton number ($L$), as can be seen by assigning ($B,L)$
charges of $(1/3,-1)$ to $\Omega$ and $(1/3,1)$ to $\chi^{-1/3}$.
The couplings $Y_{ij}'u^c_i d^c_j \chi^*$, allowed by the gauge
symmetry are forbidden by $B$, and the couplings $F_{ij}'L_i Q_j
\chi^*$, allowed by the gauge symmetry as well as $B$ are forbidden
by lepton number symmetry, which is assumed to be broken only by
soft terms.\footnote{The couplings $F_{ij}'$ of course do not
mediate rapid proton decay, however, their simultaneous presence
with the couplings of Eq. (\ref{interaction}) would lead to severe
restrictions on $F_{ij}'$, since the successful $V-A$ structure of
the SM will then be drastically altered \cite{bw}. Although not
essential, we prefer to set these $F_{ij}'$ couplings to zero by $L$
symmetry.} This breaking of $L$ by two units occurs softly via a
cubic term in the scalar potential:
\begin{equation}
V = \mu~ \Omega^\dagger H \chi^{-1/3} + {\rm h.c.}~.
\label{trilinear}
\end{equation}
The simultaneous presence of Eqs. (\ref{interaction}) and (\ref{trilinear})
would imply that neutrino masses will be generated at the loop level, as they lead to
to the effective dimension 7 operator $(Ld^c)(\overline{u^c}\,\overline{e^c}) H$ \cite{babu-leung}, once the heavy
leptoquark fields are integrated out.

The Lagrangian relevant for neutrino mass generation in component form is
\begin{eqnarray}
{\cal L}_{\nu} &=& Y_{ij} ( \nu_i d^c_j \omega^{-1/3} - e_i d_j^c
\omega^{2/3}) + F_{ij} e_i^c u_j^c \chi^{-1/3} -\mu(\omega^{-2/3}
H^+ + \omega^{1/3} H^0) \chi^{-1/3} + {\rm h.c.}
\label{Lag}
\end{eqnarray}
Once the neutral component of the SM Higgs doublet acquires a vacuum expectation value (VEV)
$v = 174$ GeV, the cubic term in the scalar potential will generate mixing
between $\omega^{-1/3}$ and $\chi^{-1/3}$ leptoquarks, with a mass matrix
given by
\begin{eqnarray}
M^2_{\rm LQ} = \left(\begin{array}{cc}m^2_\omega & \mu v \cr \mu^* v
& m^2_\chi\end{array}\right)~.
\end{eqnarray}
The parameter $\mu$ can be made real by redefining the leptoquark
fields. We diagonalize this matrix to obtain the leptoquark mass
eigenstates through
\begin{eqnarray}
\left(\begin{array}{c} \omega^{-1/3} \\ \chi^{-1/3}\end{array}
\right) = \left(\begin{array}{cc} c_\theta & s_\theta \\ -s_\theta &
c_\theta \end{array} \right) \left(\begin{array}{c} X_1^{-1/3} \\
X_2^{-1/3}\end{array}\right)
\end{eqnarray}
where $c_\theta = \cos\theta,~s_\theta = \sin\theta$ with the angle
$\theta$ given by
\begin{equation}
\tan2\theta = {2 \mu v \over m_\chi^2-m_\omega^2}~. \label{angle}
\end{equation}
The squared masses of $X_{1,2}^{-1/3}$ are given by
\begin{equation}
M^2_{1,2} = {1 \over 2} \left[m^2_\omega + m^2_\chi \mp
\sqrt{(m_\omega^2-m^2_\chi)^2 + 4 \mu^2 v^2}~  \right]
\label{eigenvalue}
\end{equation}
with $M_1^2 \leq M_2^2$.

\subsection{Two--loop neutrino mass diagrams}

\begin{figure}[t]
\centering
    \subfloat[W exchange]{{\label{w-diagram}}\includegraphics[scale=0.63]{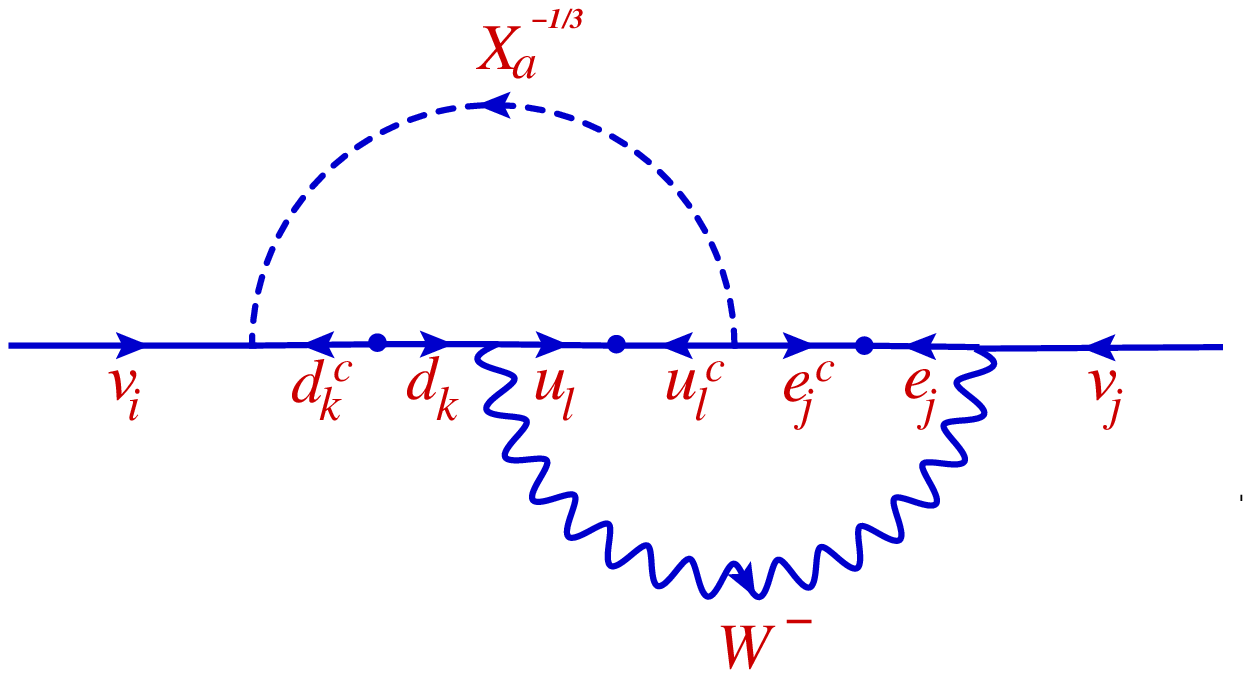}}
    \hspace{1cm}
    \subfloat[H exchange]{{\label{H-diagram}}\includegraphics[scale=0.65]{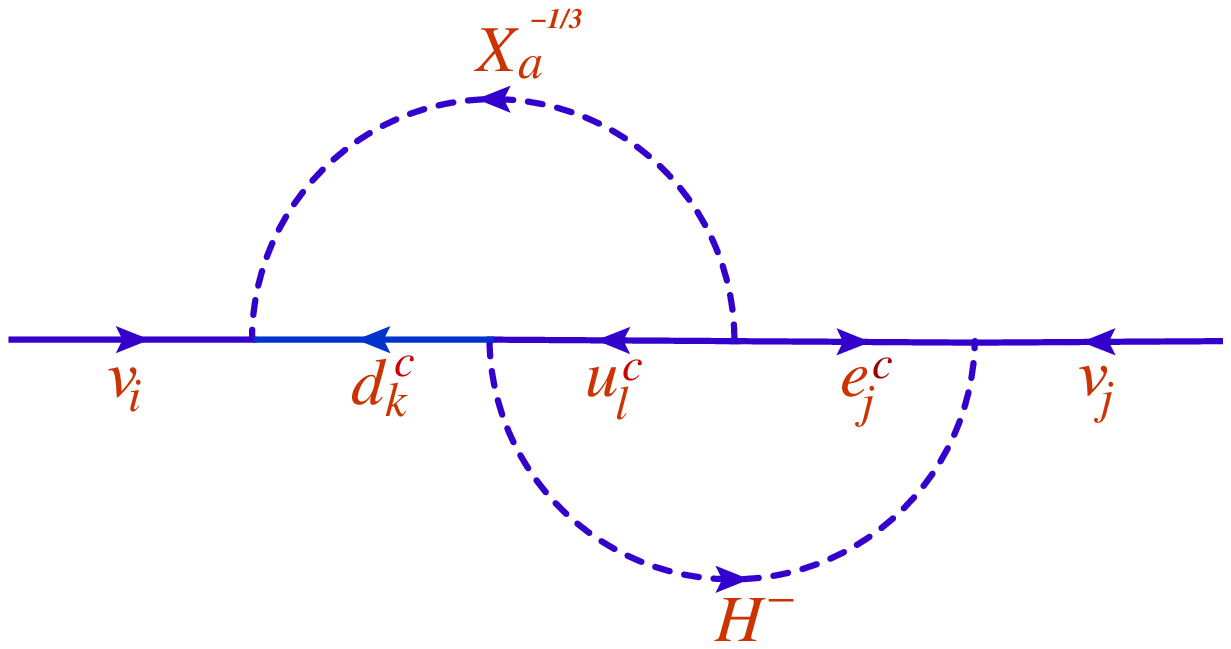}}
    \caption{\footnotesize The two--loop diagrams contributing to neutrino mass generation.}
    \label{two--loop}
\end{figure}

Since lepton number violation occurs only when all three terms of Eq. (\ref{Lag}) are
simultaneously present, neutrino masses are generated in the model only at the
two--loop level.  The relevant diagrams, which involve the exchange of leptoquarks
and a $W^\pm$ boson, are shown in
Fig. \ref{two--loop}. We have evaluated these diagrams in the Feynman
gauge. In this gauge, the unphysical charged Higgs boson exchange has to be kept.
Interestingly, this set of charged Higgs diagrams add up to zero for the neutrino mass.
The underlying reason for this result is that in the SM, both the up--type quark masses and
the down--type quark masses are generated from the same Higgs doublet.  Consequently
the charged Higgs boson Yukawa couplings to the up--type quark has a relative minus sign
compared to its Yukawa couplings to the down--type quarks, as given in the
interaction Lagrangian
\begin{equation}
{\cal L}_{H^\pm} = {1 \over v} \left[\overline{u}_R~M_u~ V ~ d_L~ H^
- - \overline{u}_L~ V~ M_d~ d_R ~H^- + {\rm h.c.} \right]
\label{charged}
\end{equation}
Here $V$ is the CKM matrix, and $M_u$ and $M_d$ are the diagonal
mass matrices for the up and down quarks. When combined with Fig.
\ref{H-diagram}, this relative minus sign implies zero net contribution
to the neutrino mass from the charged Higgs boson exchange.
In Fig. \ref{H-diagram}, the vertex $u^c_l d^c_k H^+$ is a sum of two contributions,
one where chirality flip occurs in the $u^c_l$ line, and the other where it occurs in the $d^c_i$ line.
These two contributions exactly cancel, see Eq. (\ref{charged}).

Straightforward evaluation of the
leptoquark--$W^\pm$ exchange diagrams gives the neutrino mass
matrix as:
\begin{equation}\label{Mnu0}
(M_\nu)_{ij} = \hat{m}_0 \left[Y_{ik} (D_d)_k (V^T)_{kl} (D_u)_l
(F^\dagger)_{lj} (D_\ell)_j + (D_\ell)_i (F^*)_{il} (D_u)_l
V_{lk}(D_d)_k (Y^T)_{kj}  \right] I_{jkl}~.
\end{equation}
Here $D_{u,d,\ell}$ are the (normalized) diagonal mass matrices for
up quarks, down quarks, and charged leptons:
\begin{equation}
D_u = {\rm diag.}\left[{m_u \over m_t},~{m_c \over
m_t},~1\right],~~D_d = {\rm diag.}\left[{m_d \over m_b},~{m_s \over
m_b},~1\right],~~ D_\ell = {\rm diag.}\left[{m_e \over
m_\tau},~{m_\mu \over m_\tau},~1\right]~.
\end{equation}
The overall scale $\hat{m}_0$ is given by
\begin{equation}\label{m0hat}
\hat{m}_0 = \left({C g^2 \sin2\theta \over (16 \pi^2)^2}\right)
\left({m_t m_b m_\tau  \over M_1^2}\right)~
\end{equation}
where $M_1$ is the lighter of the two charge $-1/3$ leptoquark mass and $C= 3$ is a color
factor.
\begin{figure}[t]
\centering
    \includegraphics[height=9cm]{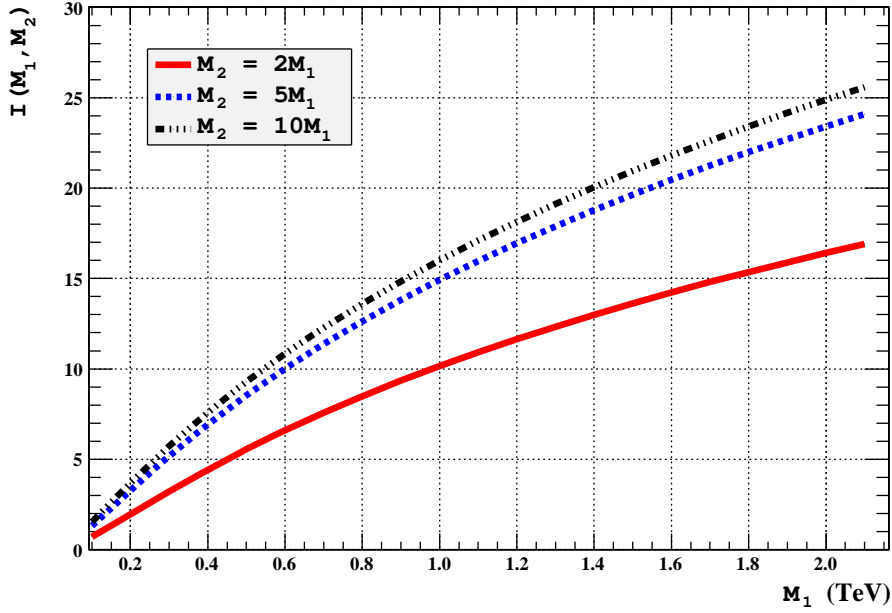}
    \caption{\footnotesize The exact loop integral with top quark inside the loop as a function of leptoquark masses
    $M_1$ and $M_2$.}
    \label{loop-integral}
\end{figure}
The dimensionless two--loop integral $I_{jkl}$ depends on the two
leptoquark masses, the $W$ boson mass, $m_W$, and the up--type quark
masses $m_{u_l}$, down--type quark masses $m_{d_k}$ and charged
lepton masses $m_{\ell_j}$, and is defined as
\begin{eqnarray}
I_{jkl}(M_1^2, M_2^2,m_W^2) &=& {M_1^2 \over m_W^2 -
m_{\ell_j}^2}\sum_{a=1,2}(-1)^a
\int_0^1 dx \int_0^\infty t dt \left({1 \over t+M_a^2}~ - ~{1 \over t+m_{d_k}^2}\right) \nonumber \\
&\times& {\rm ln} \left[{m_W^2 (1-x) + m_{u_l}^2 x + t x (1-x) \over
m_{\ell_j}^2 (1-x) + m_{u_l}^2 x + tx(1-x)}\right]~.
\end{eqnarray}
In the asymptotic limit, i.e., in the limit of $M_{1,2}^2 \gg
m_{u,d,\ell}^2,~m_W^2$, the integral $I_{jkl}(M_1^2,M_2^2,m_W^2)$
becomes independent of the flavor indices $j,k,l$ and takes the form
\begin{footnotesize}
\begin{equation}\label{asymp}
I_{jkl}(M_1^2,M_2^2,m_W^2)\simeq  \left(1- {M_1^2 \over
M_2^2}\right) \left[ 1 + {\pi^2 \over 3} + {\left(M_1^2~ {\rm ln}
{M_2^2 \over M_W^2} - M_2^2~ {\rm ln} {M_1^2 \over m_W^2} \right)
\over M_1^2 - M_2^2} + {1 \over 2}{ \left(M_1^2 ~{\rm ln}^2 {M_2^2
\over m_W^2} - M_2^2~ {\rm ln}^2 {M_1^2 \over m_W^2} \right) \over
 M_1^2 - M_2^2}\right].
\end{equation}
\end{footnotesize}
\noindent When the index $l$ is not equal to 3 (corresponding to no top quark inside the
loop), this asymptotic expression well approximates the exact integral.
The integral $I_{jk3}$ corresponding to top quark
inside the loop obtained numerically is shown in Fig. 2 as a function of $M_{1,2}$. Here we have
taken the running top quark mass $m_t(\mu = 1~{\rm TeV}) = 150$ GeV. It will turn
out that  $I_{jk3}$ will enter into the dominant source for neutrino masses.

We now proceed to write down the neutrino mass matrix explicitly. First note that if
the integral $I_{jkl}$ in Eq. (\ref{Mnu0}) is flavor universal (as
happens in the asymptotic limit as given in Eq. (\ref{asymp})), or if
$I_{jk3}$ is the only dominant contribution,
$M_\nu$ can be written as
\begin{equation}\label{Mnu1}
M_\nu = \hat{m}_0 \hat{I} \left[Y D_d V^T D_u F^\dagger D_\ell +
D_\ell F^* D_u V D_d Y^T\right] ~.
\end{equation}
where $\hat{I}$ is the flavor universal value of $I_{jkl}$.
Now, this equation can be expanded in power series in light fermion
masses.  Combining constraints from flavor changing processes, we find, to
an extremely good approximation, that the neutrino mass matrix is given by
\begin{eqnarray}\label{Mnu2}
M_\nu ~\simeq~ m_0 \left( \begin{array}{ccc} 0 & {1 \over 2} {m_\mu
\over m_\tau} x y & {1 \over 2} y \cr {1 \over 2} {m_\mu \over
m_\tau} x y & {m_\mu \over m_\tau} x z & {1 \over 2} z + {1\over 2}
{m_\mu \over m_\tau}x \cr {1 \over 2} y & {1 \over 2} z + {1 \over
2} {m_\mu \over m_\tau}x & 1+w\end{array} \right).
\end{eqnarray}
Here we have used the following definition of parameters:
\begin{eqnarray}\label{def}
x &\equiv& {F^*_{23} \over F^*_{33}}, ~~y \equiv {Y_{13} \over
Y_{33}},~~ z \equiv {Y_{23} \over Y_{33}},~ \nonumber \\
w &\equiv&
\frac{F_{32}^*}{F_{33}^*}\frac{Y_{32}}{Y_{33}}\left({m_c \over m_t}\right)\left({m_s\over m_b}\right)
\frac{I_{jk2}}{I_{jk3}} \nonumber \\
m_0 &=& 2 ~\hat{m}_0~ F^*_{33} Y_{33}~ I_{jk3}~.
\end{eqnarray}
As expected, the leading contributions are proportional to $m_t m_b m_\tau$, see
the definition of $\hat{m}_0$ in Eq. (\ref{m0hat}).  One additional contribution, denoted
as $w$ in the (3,3) entry of Eq. (\ref{Mnu2}), can be important, for a very restricted range of parameters.
This is in spite of the additional mass suppression factors $(m_c/m_t)(m_s/m_b)$ that appears in $w$, and
occurs for the parameter choice $F_{33}Y_{33} \leq 10^{-5}$ and
$M_{1,2} \sim (300 - 500) $ GeV. The Yukawa coupling combination $F_{32} Y_{32}$ appearing in $w$ is allowed
to be order one, these couplings do not lead to excessive flavor violating processes even when the
leptoquarks are light.  ($F_{32}$ couples $\tau_R$ to $c_R$, and $Y_{32}$ couples $L_3$ to $s_R$,
both of which are found to be rather un-constrained.)
Contributions not shown in Eq. (\ref{Mnu2}) have magnitudes $\ll 0.01$ eV, because
of mass suppression factors and the constraints on the Yukawa couplings arising from rare processes mediated
by the leptoquarks.

We shall consider the case $w \ll 1$ for the most part of our analysis, but we shall also examine the
special case where $w \gg 1$, which will be realized when the coupling $F_{33}$ takes anomalously small value and
the leptoquark mass is less than about 500 GeV.

The zero in the (1,1) entry of Eq. (\ref{Mnu2}) is highly
suppressed. We find, using the lowest non-vanishing contribution from Eq.
(\ref{Mnu1}) for this entry to be
\begin{equation}
(M_\nu)_{11} \simeq  y \left({F_{13}^* \over F_{33}^*}\right)
\left({m_e \over m_\tau}\right)~m_0. \label{11-entry}
\end{equation}
Now, unless $\left({F_{13}^*/F_{33}^*}\right) =
\left({F_{13}^*/F_{23}^*}\right) x$ takes values as large as
$m_\tau/m_e \sim 4000$, this entry will be negligible for neutrino
masses.  That possibility is however inconsistent with an acceptable fit
to neutrino data on the one hand, and flavor violating constraints on the
other.  To see this, let us note that a good fit to neutrino data, discussed
in more detail later in this section, requires
$|Y_{33}| \sim |Y_{23}| \sim |Y_{13}|$ and $x=|F_{23}|/|F_{33}| \sim m_\tau/m_\mu \sim 16$.
Further, the leptoquark mass should not exceed about
10 TeV, or else the neutrino mass $m_3$ will be smaller than 0.05 eV, which would be inconsistent with
the atmospheric oscillation data.  For the same reason, $F_{33} \ll 1$, while allowed, would require
the leptoquark mass to be much below a TeV.  Then flavor violation constraints become
important.  $\mu-e$ conversion in nuclei sets an upper limit on the products
$|F_{13}F_{23}|<10^{-4}$ and
$|Y_{13}Y_{23}|<10^{-4}$ corresponding to a
leptoquark mass of 300 GeV. When these limits are combined with the need to obtain the
right magnitude for $m_3$, we see that $|F_{23}|$ cannot
be smaller than $10^{-2}$.  Consequently the (1,1) entry of $M_\nu$, given in Eq. (\ref{11-entry}),
is much smaller than 0.01 eV.  This of course means that contributions to
neutrinoless double beta decay from light neutrino exchange is negligible.  See however, other
contributions to this process via the exchange of a leptoquark and
$W$ boson in Fig. \ref{double-beta} discussed later, which turns out to be significant.

The other remarkable feature of the mass matrix in Eq. (\ref{Mnu1})
is that, although it may not be obvious from its form, $M_\nu$ has
nearly zero determinant when $w \ll 1$. This would imply that one of the neutrinos
is essentially massless in the model for most of the parameter space. This can be seen by
observing that only the top quark mass, and not the $c$ and $u$
quark masses, has entered into $M_\nu$ when $w$ is set to zero.  In this case, each of the
two terms in Eq. (\ref{Mnu1}) has rank one, and the sum of the two terms has rank
two, leaving the determinant to be zero. The first non-zero entry to
the determinant in our expansion  is found to be
\begin{eqnarray}
{\rm Det}(M_\nu) &\simeq& {y \over 4} m_0^3\left({m_c \over
m_t}\right) \left({m_s \over m_b}\right) \left({m_\mu \over
m_\tau}\right)\left( \frac{I_{jk2}}{I_{jk3}}\right) \left(x{F_{32}^*
\over F_{33}^*} - {F_{22}^* \over
F_{33}^*}\right) \nonumber \\
&& \times \left[\left(y {Y_{22} \over Y_{33}}-z {Y_{12} \over
Y_{33}}\right) + x \left({m_\mu \over m_\tau}\right) \left({Y_{12}
\over Y_{33}}-y {Y_{32} \over Y_{33}}\right) \right]~.
\end{eqnarray}
Again, with $|x| \sim m_\tau/m_\mu \sim 16$, $|y|\sim1$, and $w \ll
1$ (or equivalently $Y_{32}F_{32}^* \ll 10^{4} Y_{33}F_{33}^*$), the
determinant will be much less than $(0.01~{\rm eV})^3$ and thus $m_1
\simeq 0$ to a high degree of accuracy. Note that $F_{33}^*$ cannot
be less than $10^{-3}$, or else the neutrino mass scale will be too
small.

\subsection{Predictions for neutrino oscillations when $\bm{w \ll 1}$}
We turn now to the predictions of the model when $w \ll 1$, which is
realized is much of the parameter space. Since $M_\nu$ in Eq. (\ref{Mnu2}) has only four (complex)
parameters, there are two predictions for neutrino masses and
mixings. These are summarized below:
\begin{equation}
m_1 \simeq 0,~~\tan^2\theta_{13} \simeq {m_2 \over m_3}
\sin^2\theta_{12}~.
\label{pred}
\end{equation}
Here we have used the standard parametrization of the PMNS matrix.
Furthermore, the two Majorana phases $\alpha$ and $\beta$ (the
phases of masses $m_1$ and $m_2$ respectively) are given by
\begin{equation}
\beta \simeq 2 \delta + \pi,~~ \alpha \simeq 0~,
\end{equation}
with $\alpha \simeq 0$ being a consequence of $m_1 \simeq 0$. These
predictions have been obtained in the context of textures before
(see Ref. \cite{bmt}). Here we have derived them without resorting
to textures.

To check the consistency of  these predictions, we keep
$\Delta m^2_{\rm solar} = 7.65 \times 10^{-5}$ eV$^2$, $\Delta
m^2_{\rm atm} = 2.40 \times 10^{-3}$ eV$^2$ at their central values
\cite{schwetz}, and then vary $\sin^2\theta_{12}= 0.304
^{+0.022}_{-0.016}$ in its allowed range. This gives from Eq. (\ref{pred})
\begin{equation}
\sin^2\theta_{13} = \{0.051, 0.049,~0.046,~0.044\}~,
\end{equation}
where the numbers correspond to the central value, and
$(1,~2,~3)~\sigma$ deviation in $\sin^2\theta_{12}$. This should be
compared with the limit $\sin^2\theta_{13} \leq  (0.040,~0.056)$, at
the $(2,~3) \sigma$ level \cite{schwetz}. We see broad agreement,
although $\sin^2\theta_{13}$ should be very close to the current
limit.

\begin{figure}
\centering
    \includegraphics[scale=0.7]{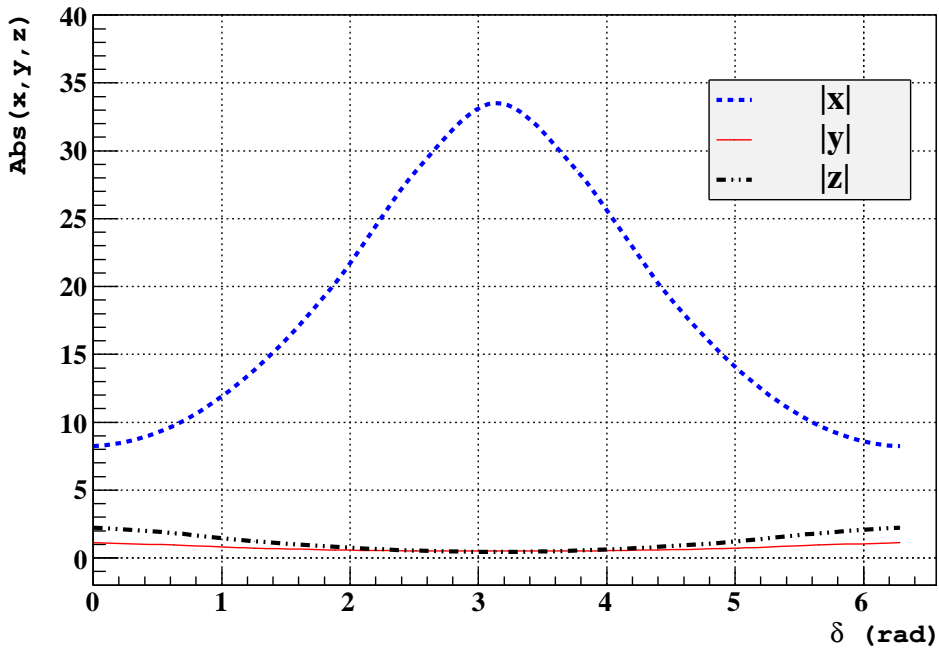}
    \caption{\footnotesize The absolute values of $x,~y,~z$}
    \label{xyz-fig}
\end{figure}

We can infer the values of $(x,~y,~z)$ in Eq. (\ref{Mnu2}) from
neutrino data.  For this purpose we use the central values from
neutrino data: $\Delta m^2_{\rm solar} = 7.65 \times 10^{-5}$
eV$^2$, $\Delta m^2_{\rm atm} = 2.40 \times 10^{-3}$ eV$^2$, along
with $\sin^2\theta_{23} = 0.50$, $\sin^2\theta_{12} = 0.304$, and
$\sin^2\theta_{13} = 0.051$ \cite{schwetz}.  We obtain
\begin{eqnarray}\label{xyz}
x = {12.21 - 4.17 e^{i\delta}  \over 0.726 + 0.248 e^{i\delta}},~~ y
= {-0.112 e^{i\delta} (2.927+e^{i\delta}) \over -0.473 +0.018 e^{i
\delta} + 0.062 e^{2i\delta}},~~~z = {0.473 + 0.342 e^{i\delta}
+0.062 e^{2i\delta} \over 0.473 -0.018 e^{i\delta} -0.062 e^{2
i\delta}}~.
\end{eqnarray}
The absolute values of $x,~y,~z$ are given in Fig. \ref{xyz-fig}. We
see that  $|x|\gg1$ and $|y|,~|z|
\sim 1$.  These values will directly affect the branching ratios
of the leptoquarks, making a strong connection between collider
physics and neutrino physics.

\subsection{A special case with $\bm{w \gg 1}$}

It is interesting to see the constraints on the model when $w \gg 1$. This
occurs for a small range of parameters when the leptoquark mass is less
than about 500 GeV, and when $|F_{33} Y_{33}| \ll 10^{-5}$.
The latter condition can be satisfied when either $|F_{33}|$
or $|Y_{33}|$ is much smaller than $10^{-5}$. Both of these couplings cannot
be simultaneously very small, or else the (2,3) entry of $M_\nu$ will
become unacceptably small.



When $F_{33} \rightarrow 0$, we have the condition $(M_\nu)_{13} \simeq 0$, see Eq. (\ref{Mnu2}). Since
$(M_\nu)_{11} \simeq 0$ in this case we reproduce a well known texture model
\cite{marfatia,xing}. This limit of our model leads to the following predictions for
$\sin^2\theta_{13}$ and the phase parameters:
\begin{eqnarray}
\sin^2\theta_{13} &\simeq& \frac{|\Delta m_{\rm solar}^2|/|\Delta
m_{\rm atm}^2|}
{\left|\cot^2\theta_{12}-e^{i2(\alpha-\beta)}\tan^2\theta_{12}\right|},
\\
\tan 2(\beta+\delta) &\simeq& \frac{\tan^2\theta_{12} \sin
2(\alpha-\beta)}{\cot^2\theta_{12}-\tan^2\theta_{12} \cos 2
(\alpha-\beta)}.
\end{eqnarray}
This gives $0.012 \leq \sin^2\theta_{13} \leq 0.017$, which is
within reach of the next generation long baseline neutrino
oscillation experiments.  This case requires the leptoquarks to be
lighter than about 500 GeV, which are within reach of the LHC.

\subsection{The case of $\bm{w \sim 1$}} \label{w3}

For the case when $w \sim 1$, there are no restrictions on the neutrino oscillation parameters,
except the one arising from the vanishing of the (1,1) entry of $M_\nu$.
This relation can be taken to be a prediction for $m_1$, and an additional relation for
the phase parameters.  These are given by
\begin{eqnarray}
\tan^2\theta_{13} &\simeq& \frac{m_2}{m_3} \sin^2\theta_{12} \left|1
+ \frac{m_1}{m_2} \cot^2\theta_{12}e^{i(\alpha-\beta)} \right|,
\nonumber \\
2\delta &\simeq& \beta - \pi - \cot^{-1} \left[ \frac{1 + a
\cos(\beta-\alpha)}{a\sin(\beta-\alpha)} \right],
\end{eqnarray}
where $a\equiv \tfrac{m_1}{m_2}\cot^2\theta_{12}$. In this scenario $\theta_{13}$
can take any value between zero and its current experimental upper limit.
For $w \sim 1$, the leptoquark mass cannot exceed about 500 GeV, so this scenario
is testable at the LHC.

\subsection{Limit on the parameter $\bm{\mu}$}

There are certain restrictions on the parameter $\mu$ that enters in
the neutrino mass Lagrangian of Eq. (\ref{Lag}). It is this term
that is responsible for the mixing of $\om^{-1/3}$ with
$\chi^{-1/3}$.  Since the $SU(2)_L$ partner of $\omega^{-1/3}$ does
not mix with any other field, the $\mu$ term will induce new
contributions to the electroweak $\rho$ parameter (or equivalently
the $T$ parameter). This is because mixing via the $\mu$ term splits
the masses of the $SU(2)_L$ doublet members. This mass-splitting
must obey \cite{pdg}
\begin{equation}
\frac{C}{3}
(\Delta M)^2 \leq (57\,{\rm GeV})^2,
\label{splitting}
\end{equation}
where $C=3$ for leptoquarks.
In our model, apart from the $\mu$ term, there is a quartic scalar coupling,
$V \supset \kappa |(\Omega^\dagger H)|^2$, which contributes to the $\omega^{-1/3}$ mass,
and not to the $\omega^{2/3}$ mass, causing a further splitting.
The parameter $(\Delta M)^2$ in Eq. (\ref{splitting}) is given by
\begin{equation}
\Delta M \simeq \frac{\kappa v^2 + M_1^2 - m_\omega^2}{2M_1}~,
\label{rho}
\end{equation}
where we have assumed that the two states are nearly degenerate. The
two--loop induced neutrino mass is maximized when the mixing
parameter $\sin2\theta = 1$.  For this choice, one has the relations
$M_{1,2}^2 = m_\omega^2 \mp \mu v$.  When the leptoquark mass is
well above 200 GeV, the contribution $\kappa v^2$ in Eq.
(\ref{rho}), which is at most of order (200 GeV)$^2$, can be
neglected in relation to the $\mu v$ term.  In this case we obtain
from Eq. (\ref{splitting}) an upper limit on $\mu$:  $\mu \leq 0.65
M_1$.  We shall use this constraint in deriving an upper limit on
the leptoquark mass.

There are other reasons why $\mu$ cannot be arbitrarily large: the
positivity of leptoquark squared mass (needed for color
conservation) and the perturbativity of the theory. The first of
these conditions implies, from Eq. (\ref{eigenvalue}), that
\begin{equation}
\mu \leq \frac{m_\om m_\chi}{v}. \label{mu-1}
\end{equation}

\begin{figure}[t]
\centering
    \includegraphics[scale=0.5]{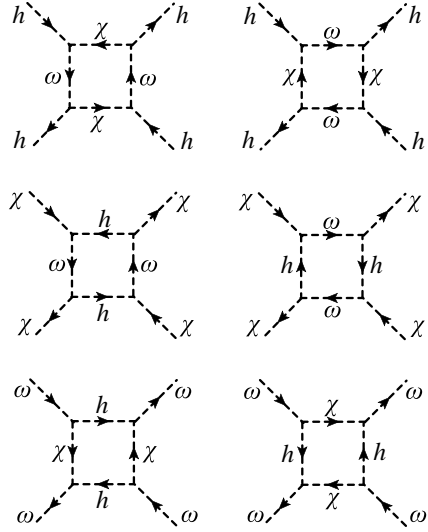}
    \caption{\footnotesize The diagrams leading to the quartic couplings
    correction generated by $\mu$.}
    \label{box}
\end{figure}

The second constraint emerges because $\mu$ induces negative quartic
couplings for the three fields $\chi^{1/3}$, $\omega^{1/3}$ and
$H^0$. These induced couplings cannot exceed the corresponding
tree--level quartic couplings which can at most be of order one.
Otherwise the theory would be non-perturbative, or electric charge
and color would break in the minimum of the theory. We parameterize
the effective quartic couplings of the three fields as
\begin{equation}
-\mathcal{L}_{\rm eff} = \lambda_{\rm eff}
\left(H^{0*}H^0\right)^2+\lambda'_{\rm eff}
\left(\chi^*\chi\right)^2+ \lambda''_{\rm eff}\left(\omega^*\omega
\right)^2,
\end{equation}
with $H^0=h/\sqrt{2}$. The 1--loop correction to these quartic
couplings are shown in Fig. \ref{box}.\footnote{For the $h^4$ term,
there is a similar diagram generated by the SM $(H^\dagger H)^2$
quartic coupling. But
here we focus on the diagrams generated by  the $\mu$ term.} By
evaluating these diagrams we obtains for the effective couplings,
\begin{eqnarray}
\lambda_{\rm eff} &=& -\frac{3}{32\pi^2}
\frac{\mu^4}{(m_\chi^2-m_\omega^2)^2}
\left[\frac{\mc+\mo}{\mc-\mo}~{\rm ln}\frac{\mc}{\mo}-2 \right]
\nonumber \\
\lambda'_{\rm eff} &=&
-\frac{1}{128\pi^2}\frac{\mu^4}{(m_\chi^2-m_h^2)^2}
\left[\frac{\mc+m_h^2}{\mc-m_h^2}~{\rm ln}\frac{\mc}{m_h^2}-2
\right]
\nonumber \\
\lambda''_{\rm eff} &=&
-\frac{1}{128\pi^2}\frac{\mu^4}{(\mo-m_h^2)^2}
\left[\frac{\mo+m_h^2}{\mo-m_h^2}~{\rm ln}\frac{\mo}{m_h^2}-2
\right], \label{lambda-eff}
\end{eqnarray}
where the factor 3 in $\lambda_{\rm eff}$ is a color factor. Since
these effective couplings are negative, they must be smaller in
magnitude compared to the corresponding tree level ones, otherwise
the potential would become unbounded, suggesting instability of the
vacuum. By demanding all these induced couplings to be less than
one, so that the theory is perturbative with a stable vacuum, one
can derive limits of $\mu$. Of course, if the limit derived from Eq.
(\ref{mu-1}) turns out to be more stringent, that will supersede the
present limits.

For illustration, we consider a few simplified cases.
\begin{itemize}
\item $m_h\approx m_\om \ll m_\chi$. For $m_h=m_\om=200~{\rm GeV}$ and $m_\chi=1~{\rm TeV}$, the bound
from Eq. (\ref{mu-1}) gives a more stringent limit, i.e. $\mu<1.1~{\rm
TeV}$. For $m_\chi=2~{\rm TeV}$ and the other masses the
same as above, the more stringent limit comes from the $\lambda_{\rm eff}$, viz.,
$\mu<1.9~{\rm TeV}$.

\item $m_h \ll m_\om \neq m_\chi$: In this case
\begin{equation}
\mu < m\left(\frac{64\pi^2}{{\rm ln}\frac{m}{m_h}-1}\right)^{1/4},
\end{equation}
where $m$ is the smallest of $m_\om$ and $m_\chi$. For example, if
$m_\om=1~{\rm TeV}$, $m_\chi=2~{\rm TeV}$, and $m_h=200~{\rm GeV}$,
$\mu<5.7~{\rm TeV}$.
\item $m_h \ll m_\om \approx m_\chi$: In this case
\begin{equation}
\mu < m_\chi(64\pi^2/\xi)^{1/4},
\end{equation}
where $\xi = {\rm max}~\{1,~{\rm ln}\tfrac{m_\chi}{m_h}-1\}$. The
choice of $\xi$ depends upon the value of $m_\om$ and $m_\chi$. For
example, if $m_\om=m_\chi=1~{\rm TeV}$ and $m_h=200~{\rm GeV}$, one
uses $\xi=1$, so $\mu<5~{\rm TeV}$. On the other hand, if
$m_\om=m_\chi=2~{\rm TeV}$ and $m_h=200~{\rm GeV}$, one uses
$\xi=\left({\rm ln}\tfrac{m_\chi}{m_h}-1\right)$, so $\mu<9.4~{\rm
TeV}$.
\end{itemize}

\subsection{Limit on leptoquark masses}

The scaling of the neutrino mass with the LQ mass goes as $m_\nu
\sim (m_t m_b m_\tau \mu v)/M_1^4$. Since our neutrino mass matrix
has a normal hierarchy structure, the neutrino mass scale $m_0$
should be around 0.02 eV. We shall assume that all the Yukawa
couplings are bounded by 1, a requirement for perturbative treatment
of the problem.  As we have seen, a fit to the neutrino oscillation
data requires $F_{33}/F_{23} \sim m_\mu/m_\tau \sim 1/16$.  The
maximum value of $m_0$ is realized for $\sin2\theta = 1$,
corresponding to largest allowed LQ mass.  From these considerations
we arrive at a limit on the LQ mass of 12 TeV.  If the leptoquarks
are in the multi-TeV range, all three members should be nearly
degenerate, from the requirement of $\sin2\theta \approx 1$ and the
$\rho$ parameter constraint. A slightly better limit can be derived
on their masses from the rare process $\mu-e$ conversion in nuclei
(see the next section for details).  From the experimental limit on
this process, one can derive the upper limit on $|Y_{13}^*Y_{23}|$
as a function of $\omega^{2/3}$ mass. Since neutrino mass fitting
requires $Y_{i3}$  to be of the same order for $i=1-3$, one can also
determine an upper limit on $Y_{33}$. Combining all of these, we
obtain an upper limit of 10 TeV on $M_1$.

There are lower limits on leptoquark mass from  the Tevatron where
they could be pair produced. The D{\O} and CDF experiments
\cite{fermilab} have obtained limits on the second and the third
generation leptoquarks of 316 GeV and 245 GeV respectively
\cite{fermilab}. Since those experiments look for b-jets and missing
energy, these bounds are strictly applicable for leptoquarks coupled
to neutrinos. In our case, $\om^{-1/3}$ and $\chi^{-1/3}$ have
direct couplings to the neutrinos (the latter from mixing) for which
the quoted limits apply.  As for $\om^{2/3}$, being a member of the
same $SU(2)_L$ doublet as $\omega^{-1/3}$, it should be nearly
degenerate with $\omega^{-1/3}$, and thus the limits of Ref.
\cite{fermilab} should apply to $\omega^{2/3}$ as well.

\section{Constraints from rare processes}

Since this model features lepton flavor (as well as total lepton
number) violation, it is important to see how its parameters are
constrained by the experimental data, especially those arising from
rare decay processes that are forbidden in the SM. Here we derive a
variety of limits on the couplings $Y_{ij}$ and $F_{ij}$ as
functions of the LQ masses.  These limits should  be (and have been)
satisfied in the neutrino fits.  The processes we shall consider are
$\mu^- \to e^- \gamma$, $\mu^- \to e^+e^-e^-$, $\mu-e$ conversion in
nuclei, $\tau^- \to e^-\eta$, $\tau^- \to \mu^- \eta$,
$B_{s,d}-\ol{B}_{s,d}$ mixing, $K-\ol{K}$ mixing, $D-\ol{D}$ mixing,
$D_s^\pm \rightarrow \ell^\pm \nu$ decay, muon $g-2$, $\pi^+\to
\mu^+ \ol{\nu_e}$ decay, and neutrinoless double beta decay.  The
expected rates for several of these processes are found to be in the
interesting range for the next generation of experiments.

\subsection{$\bm{\mu^- \to e^- \gamma}$}

\begin{figure}[t]
\centering
    \includegraphics[scale=0.85]{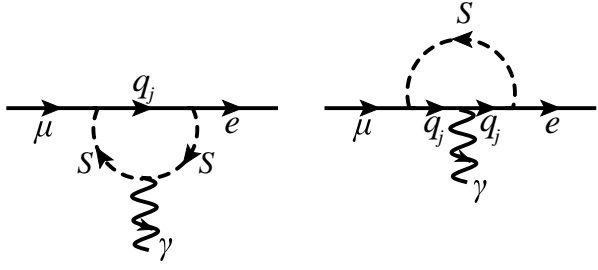}
    \caption{\footnotesize The diagrams leading to $\mu\to e
    \gamma$. $q_j$ and $S$ represent $d_j$ ($u_j^c$) quarks and $\om^{2/3}$
    ($\chi^{1/3}$) respectively.}
    \label{mu-egamma}
\end{figure}

The rare decay $\mu^- \to e^- \gamma$ arises from a one--loop diagram. In
the limit of $m_e=0$, there are only three diagrams for this decay shown in Fig.
\ref{mu-egamma}. In general, the amplitude for this decay process can
be written as \cite{kuno}
\begin{eqnarray}
\mathcal{M} &=&
-e\frac{iq_\nu}{m_\mu}\ol{u}(p_e)\sigma^{\mu\nu}(f_{M1}+f_{E1}\gamma_5)u(p_\mu)\epsilon_\mu(q)
\end{eqnarray}
where $m_\mu$ is the muon mass, $q\equiv p_\mu-p_e$ is the photon
momentum, and $\epsilon_\mu$ is the photon polarization vector. The
effective couplings $f_{M1}$ and $f_{E1}$ are found to be (repeated
indices implies summation)
\begin{eqnarray}
f_{E1} &=& \frac{3m_\mu^2}{32\pi^2}\left(
\frac{Y^*_{1j}Y_{2j}}{m_\om^2}F_3(x_{d_j})
-\frac{F_{1j}F^*_{2j}}{m_\chi^2}F_4(x_{u_j})\right),\nonumber \\
f_{M1} &=& \frac{3m_\mu^2}{32\pi^2} \left(
\frac{Y^*_{1j}Y_{2j}}{m_\om^2}F_3(x_{d_j})
+\frac{F_{1j}F^*_{2j}}{m_\chi^2}F_4(x_{u_j})\right),
\end{eqnarray}
where $x_{d_j}\equiv m_{d_j}^2/m_\om^2$, $x_{u_j}\equiv
m_{u_j}^2/m_\chi^2$; $m_{u_j}$ and $m_{d_j}$ are the $j$-th
generation of up-- and down--type quark masses respectively; $m_\om$
and $m_\chi$ are $\om^{2/3}$ and $\chi^{1/3}$ masses respectively;
and $F_3(x)$ and $F_4(x)$ are the dimensionless functions given by
\cite{lavoura,hisano}
\begin{eqnarray} F_3(x) &=&
-\frac{x}{12}~\frac{(1-x)(5+x)+2(2~x+1)~{\rm
ln}~x}{(1-x)^4}, \nonumber \\
F_4(x) &=& -\frac{1}{12}~\frac{(1-x)(5~x+1)+2~x(2+x)~{\rm
ln}~x}{(1-x)^4}.
\end{eqnarray}
These expressions assume that the leptoquark mixing angle
$\theta$ (see Eq. (\ref{angle})) is zero. For non-zero $\theta$,
one can make the following replacement:
\begin{eqnarray}
\frac{1}{m_\chi^2} \to \sum^2_{a=1}\frac{\zeta_a}{M_a^2},
\end{eqnarray}
where $\zeta_1=\sin^2\theta$, $\zeta_2=\cos^2\theta$, and $M_a$ are
defined in Eq. (\ref{eigenvalue}).

It is an excellent approximation to set the first and second generation
quark masses to zero. The branching ratio for this decay is
found to be
\begin{equation}
{\rm BR} (\mu^-\ \to e^- \gamma) = \frac{27\alpha}{16\pi
G_F^2}\left(
\frac{\left|F_3(x_{b})Y_{13}^*Y_{23}\right|^2}{m_\om^4}+
\frac{\left|\tfrac{1}{12}F_{11}F_{21}^*+\tfrac{1}{12}F_{12}F_{22}^*+F_4(x_{t})F_{13}^*F_{23}\right|^2}{m_\chi^4}
 \right),
\label{mu-e-gamma}
\end{equation}
where $\alpha=e^2/4\pi$ is the fine structure constant.
Constraints from other $\ell_i\to \ell_j\gamma$ processes are presented in Table
\ref{rare}, where we have used the approximation $m_{\ell_j}=0$.

An interesting feature of this analysis is that the $Y_{ij}$
couplings are only weakly constrained from these processes.  This is
owing to a GIM--like cancelation for the amplitude of this process
from the two diagrams (Fig. {\ref{mu-egamma}). In the limit of
down-type quark mass being zero, the two graphs exactly cancel. This
cancelation occurs because the charge of the internal leptoquark
($2/3$) is twice as large and opposite in sign compared to the
charge of of the internal down quark ($-1/3)$.  The amplitude of the
graph where the photon is emitted from the scalar line is twice
smaller compared to the one where the photon is emitted from the
fermion line, leading to the cancelation.  The non-canceling
contribution is suppressed by a factor $m_b^2/m_\omega^2$ in the
amplitude, which weakens the limit.

\begin{table}[t]
\begin{center}
\begin{tabular}{ccc}\hline\hline
Process & BR & Constraint \\ \hline \\ $\mu \to e \gamma$ & $< 1.2
\times 10^{-11}$ & $\frac{\left|F_3(x_{b})Y_{13}^*Y_{23}
\right|^2}{m_\omega^4} +
\frac{\left|\frac{1}{12}F_{11}F_{21}^*+\frac{1}{12}F_{12}F_{22}^*+F_4(x_{t})F_{13}F_{23}^*\right|^2}{m_\chi^4}
< \frac{3.1\times 10^{-19}}{{\rm GeV^4}} $
\\\\
$\tau \to e \gamma$ & $< 1.1 \times 10^{-7}$ &
$\frac{\left|F_3(x_b)Y_{13}^*Y_{33} \right|^2}{m_\om^4} +
\frac{\left|\frac{1}{12}F_{11}F_{31}^*+\frac{1}{12}F_{12}F_{32}^*+F_4(x_{t})F_{13}F_{33}^*\right|^2}{m_\chi^4}
< \frac{1.6\times 10^{-14}}{{\rm GeV^4}}$
\\\\
$\tau \to \mu \gamma$ & $< 4.5 \times 10^{-8}$ &
$\frac{\left|F_3(x_{b})Y_{23}^*Y_{33} \right|^2}{m_\om^4} +
\frac{\left|\frac{1}{12}F_{21}F_{31}^*+\frac{1}{12}F_{22}F_{32}^*+F_4(x_{t})F_{23}F_{33}^*\right|^2}{m_\chi^4}
< \frac{6.7\times 10^{-15}}{{\rm GeV^4}}$
\\\\
\hline \hline
\end{tabular}
\caption{{\footnotesize The constraints from $\ell_i \to \ell_j
\gamma$.}} \label{rare}
\end{center}
\end{table}

\subsection{$\bm{\mu^- \to e^+e^-e^-}$} 

These  processes also occur at the one--loop level in our model (see Fig.
\ref{mu-3e}). Here, there exist photon, $Z$--boson, and the SM Higgs--boson
exchange diagrams, as well as box
diagrams. The SM Higgs diagrams are suppressed by the
electron mass, and thus highly suppressed and therefore ignored.
Therefore, only the photon,
$Z$--boson, and the box diagrams will be evaluated.

In the case of photon exchange, because the photon is now off-shell, and
with the electron mass set to zero, the amplitude is given by
\begin{eqnarray} \mathcal{M}_\gamma &=&
\frac{e^2}{m_\mu^2}\left[\ol{u}(p_{e1})\gamma^\mu(f_{E0}+f_{M0}\gamma_5)u(p_\mu)\right]\left[
\ol{u}(p_{e2})\gamma_\mu v(p_{e3})\right] \nonumber \\
&&+~ e^2 \frac{iq_\nu}{m_\mu q^2}
\left[\ol{u}(p_{e1})\sigma^{\mu\nu}(f_{M1}+f_{E1}\gamma_5)u(p_\mu)\right]
\left[ \ol{u}(p_{e2})\gamma_\mu v(p_{e3})\right] \nonumber \\
&& - \left(p_{e1}\leftrightarrow p_{e2}\right),
\end{eqnarray}
where $q\equiv p_\mu-p_{e1}$. $p_\mu$ and $p_{ei}$ are the incoming
muon and outgoing electron momenta respectively. The couplings
$f_{E0}$ and $f_{M0}$ are found to be
\begin{eqnarray}
f_{E0} &=& \frac{3m_\mu^2}{32\pi^2} \left(
\frac{Y_{1j}^*Y_{2j}}{m_\om^2}g_j +
\frac{F_{1j}F^*_{2j}}{m_\chi^2}h_j\right),
\nonumber \\
f_{M0} &=& -\frac{3m_\mu^2}{32\pi^2} \left(
\frac{Y_{1j}^*Y_{2j}}{m_\om^2}g_j - \frac{F_{1j}F^*_{2j}}{m_\chi^2}
h_j\right),
\end{eqnarray}
where $g_{1,2}=\tfrac{1}{27}(2-3\ln \tfrac{q^2}{m_\om^2})$,
$h_{1,2}=\tfrac{1}{54}(5-12\ln \tfrac{q^2}{m_\chi^2})$,
$g_3=F_1(x_b)$, and $h_3=F_2(x_t)$. The dimensionless functions
$F_1(x)$ and $F_2(x)$ are given by \cite{hisano}
\begin{eqnarray}
F_1(x) &=& \frac{-4+9x-5x^3+2(2x^3+3x-2)\ln x}{36(1-x)^4}, \nonumber \\
F_2(x) &=&\frac{(x-1)(10+x(x-17))+2(x^3+6x-4)\ln x}{36(1-x)^4}.
\end{eqnarray}
\begin{figure}[t!]
\centering
    \includegraphics[scale=0.7]{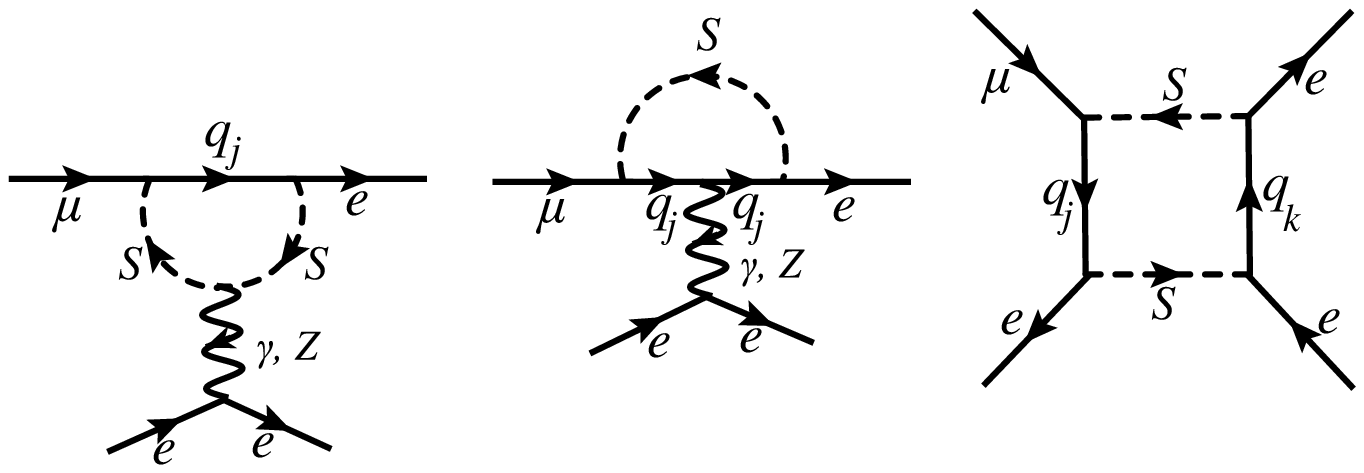}
    \caption{\footnotesize The diagrams leading to $\mu\to 3e$. $q_j$ and $S$ represent $d_j$ ($u_j^c$) quarks and $\om^{2/3}$
    ($\chi^{1/3}$) respectively.}
    \label{mu-3e}
\end{figure}

For the $Z$ boson exchange diagram, the leading contribution after ignoring
terms suppressed by $m_\mu^2/M^2$, with $M$ being leptoquarks mass,
is
\begin{eqnarray}
\mathcal{M}_Z &=& \frac{G_F}{\sqrt{2}} \left[\eta_L
\ol{u_L}(p_{e1})\gamma^\mu u_L(p_\mu) + \eta_R
\ol{u_R}(p_{e1})\gamma^\mu u_R(p_\mu)\right] \nonumber \\
&& \times
\left[\ol{u}(p_{e2})\gamma_\mu(g_V^e-g_A^e\gamma_5)v(p_{e3})\right]
\nonumber \\
&&-\left( p_{e1}\leftrightarrow p_{e2} \right),
\end{eqnarray}
where $g_V^e=-\tfrac{1}{4}+\sw$ and $g_A^e=-\tfrac{1}{4}$. Since we
set the first and second generation quark masses to zero (which is a good
approximation), the
effective couplings, $\eta_L$ and $\eta_R$, are given by
\begin{eqnarray}
\eta_L &=& \frac{3}{2\pi^2}Y^*_{13}Y_{23}F_5(x_{b}), \nonumber \\
\eta_R &=& \frac{3}{2\pi^2}F_{13}F^*_{23}F_5(x_t),
\end{eqnarray}
with
\begin{equation}
F_5(x) = -\frac{x}{2}~\frac{1-x+\ln x}{(1-x)^2}.
\end{equation}

For the box diagrams, the amplitude can be written as
\begin{eqnarray}
\mathcal{M}_{\rm box} &=& \frac{3}{64\pi^2m_\om^2}
\left(YY^\dagger\right)_{11}\left(YY^\dagger\right)_{21}\left[\ol{u_L}(p_{e1})\gamma^\mu
u_L(p_\mu)\right]
\left[\ol{u_L}(p_{e2})\gamma_\mu v_L(p_{e3}) \right] \nonumber \\
&& +
\frac{3}{64\pi^2m_\chi^2}\left(FF^\dagger\right)_{11}\left(FF^\dagger\right)_{12}\left[\ol{u_R}(p_{e1})\gamma^\mu
u_R(p_\mu)\right]
\left[\ol{u_R}(p_{e2})\gamma_\mu v_R(p_{e3}) \right] \nonumber \\
&& - \left(p_{e1}\leftrightarrow p_{e2} \right).
\end{eqnarray}
Here we have set all the quark masses to zero.

We now proceed to the calculation of the total amplitude.
In the loop integral functions, $F_1...F_5$, since the
$Y_{ij}$ couplings are not
constrained by $\mu^- \to e^- \gamma$, one needs to include all the
diagrams mediated by $\om^{2/3}$ boson, since the $Y_{ij}$ couplings may be of
order one. However, since the $Z$--boson exchange contribution is
suppressed by the the bottom quark mass, it can safely be ignored. On the
other hand, the $Z$ exchange is significant for $\chi^{1/3}$
mediated process with the top quark inside the loop. The electron
mass has to be included in the calculation in order to avoid
infrared singularity. The branching ratio is found to be
\begin{eqnarray}
{\rm BR} (\mu^- \to e^+e^-e^-) &=&
\left(\frac{3\sqrt{2}}{32\pi^2G_F}\right)^2~
\left(\frac{C^L_{jk}Y_{1j}^*Y_{2j}Y_{1k}Y_{2k}^*}{m_\om^4} +
\frac{C^R_{jk}F_{1j}^*F_{2j}F_{1k}F_{2k}^*}{m_\chi^4} \right),
\end{eqnarray}
where we have introduced Hermitian parameters $C^{L,R}_{jk}$, given
by
\begin{eqnarray}
C^L_{jk} &=& \left\{ \begin{array}{l} \frac{1}{7776}\left[ 72
e^4\ln^2 \frac{m_\mu^2}{m_\om^2} -108 \left(3e^4+2e^2\yyd\right)\ln
\frac{m_\mu^2}{m_\om^2} \right.
\nonumber \\
\left. + \left(449+68\pi^2\right)e^4 +
486e^2\yyd +243 \yyd^2\right], ~{\rm for}~j,k=1,2; \\ \\
\frac{1}{288} \left[ 54e^4 F_1(x_b)-4e^2\left(
6e^2F_1(x_b)+\yyd\right) \left(\ln \frac{m_\mu^2}{m_\om^2} + i\pi
\right) +36e^2\yyd F_1(x_b) \right. \nonumber
\\
+ \left. 9e^2\yyd +9 \yyd^2\right],~ {\rm for},~j=1,2 ~ {\rm and}~
k=3; \\ \\
\frac{1}{32} \left[ 24 e^4F_1^2(x_b)+8e^2\yyd F_1(x_b) + \yyd^2
\right], ~ {\rm for}~j=k=3;
\end{array}
\right. \\ \\
C^R_{jk} &=& \left\{ \begin{array}{l} \frac{e^4}{7776} \left(
288\ln^2 \frac{m_\mu^2}{m_\chi^2} -1584 \ln \frac{m_\mu^2}{m_\chi^2}
+
108 \ln \frac{m_\mu^2}{m_e^2} + 272 \pi^2 +2759 \right) \nonumber \\
+\frac{e^2}{72}\ffd\left(11-4\ln \frac{m_\mu^2}{m_\chi^2}\right) +
\frac{1}{32}\ffd^2,~{\rm
for}~j,k=1,2; \\ \\
\frac{e^4}{24} \left[
11F_2(x_t)-29F_4(x_t)-4F_4(x_t)\ln\frac{m_\mu^2}{m_e^2}-4\left(F_2(x_t)-2F_4(x_t)\right)\left(
\ln \frac{m_\mu^2}{m_\chi^2} +i\pi\right) \right] \nonumber \\
+\frac{e^2}{144}\left[
4\sqrt{2}(6\sw-1)G_Fm_\chi^2F_5(x_t)\left(-4\ln
\frac{m_\mu^2}{m_\chi^2} - i 4\pi +11
\right) \right. \nonumber \\
+  \left.\ffd \left(18F_2(x_t)-36F_4(x_t)-4\ln
\frac{m_\mu^2}{m_\chi^2} -i4\pi+11
\right) \right]\nonumber \\
+ \frac{1}{32}\ffd \left[16\sqrt{2}\sw
G_Fm_\chi^2F_5(x_t)+\ffd\right],~{\rm for}~j=1,2~{\rm and}~k=3; \\ \\
e^4\left[\frac{3}{4}F_2^2(x_t)-3F_2(x_t)F_4(x_t)+2F_4^2(x_t)\left(
\ln \frac{m_\mu^2}{m_e^2} -\frac{11}{4}\right) \right] \nonumber \\
+\frac{e^2}{4}\left[F_2(x_t)-2F_4(x_t)\right]\left[4\sqrt{2}(6\sw-1)G_Fm_\chi^2F_5(x_t)+\ffd\right]
\nonumber \\
+2\left(1-4\sw+12\sin^4\theta_W \right)G_F^2m_\chi^4F_5^2(x_t)
\nonumber \\
+ \sqrt{2}\ffd\sw G_F m_\chi^2F_5(x_t)+ \frac{1}{32}\ffd^2, ~{\rm
for}~j=k=3.
\end{array}
\right.
\end{eqnarray}
For heavy quarks inside the loop, the factors
$C^{L,R}_{33}$ are in the agreement with \cite{hisano}.

Regarding the neutrino oscillation data, one needs the limit on the
coupling constants that appear in Eq. (\ref{def}). From the
branching ratio above, in the assumption that there is no accidental
cancelation among the various contributions, so that one can omit
terms like $Y_{13}^*Y_{23}Y_{jk}$ or $F_{13}F_{23}^*F_{jk}$ with
$j,k=1,~2$, one gets, for a LQ mass of 1 TeV,
\begin{equation}
 |Y_{13} Y_{23}| < 7.6 \times 10^{-3},\,~~~
|F_{13}F_{23}^*| < 1.8 \times 10^{-3}
\end{equation}
if $|F_{13}| \ll e$ and
\begin{equation}
|F_{13}F_{23}| < 1.3 \times 10^{-3}
\end{equation}
if $|F_{13}|\sim 1$.


\subsection{$\bm{\mu-e}$ Conversion in the nuclei}

\begin{figure}[t!]
\centering
    \includegraphics[scale=0.9]{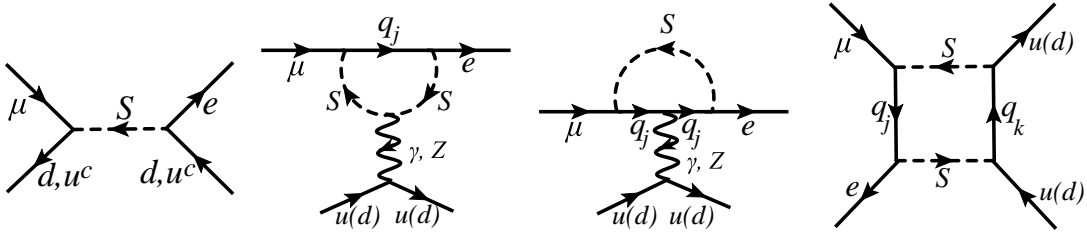}
    \caption{\footnotesize The diagrams leading to $\mu-e$ conversion.
    $q_j$ and $S$ represent $d_j$ ($u_j^c$) quarks and $\om^{2/3}$
    ($\chi^{1/3}$) respectively.}
    \label{me-conversion}
\end{figure}
Another lepton flavor violating process that may occur in this model
is $\mu-e$ conversion in  nuclei ($\mu N \to e N$). The
interesting type is the so-called coherent conversion in which the
nucleus does not change its initial state \cite{marciano}. In this
section we shall discuss the implications of our model in such
processes. It turns out that there exist tree level diagrams in
addition to the one--loop diagrams (Fig. \ref{me-conversion}).
These two sets of diagrams probe different Yukawa couplings.  The
loop diagrams involve the same couplings that appear in the neutrino
mass matrix, and has to be given special attention.
These loop
diagrams are similar to $\mu^- \to e^+e^-e^-$ except the pair of
up/down quark is attached to the photon and $Z$ lines. Following
\cite{kuno}, the branching ratio of this process is given by
\begin{eqnarray}
{\rm BR}(\mu N \to e N) &=&
\frac{|\vec{p}_e|E_em_\mu^3G_F^2\alpha^3Z_{\rm
eff}^4F_p^2}{8\pi^2Z~\Gamma_N}
\left|g_L^u(A+Z)+g_L^d(2A-Z)+2Z\Delta g_L \right|^2 \nonumber \\
&& + L \leftrightarrow R,
\end{eqnarray}
where $|\vec{p_e}|$ and $E_e$ are the momentum and energy of the
outgoing electron respectively, $Z$ is the atomic number of the
nucleus, $A$ is the mass number of the nucleus, $Z_{\rm eff}$ is the
effective atomic number defined in \cite{chiang}, $F_p$ is the
nuclear matrix element defined in \cite{kuno,marciano,chiang},
$\Gamma_N$ is the muon capture rate of nucleus $N$, and
$g^{u,d}_{L,R}$ are defined as
\begin{eqnarray}
g^u_{L} &=& \eta_{L}\left(\frac{1}{4}-Q_u\sw\right),~g^d_{R} =
\eta_{R}\left(\frac{1}{4}-Q_d\sw\right),
\nonumber \\
g^d_L &=& \eta_{L}\left(\frac{1}{4}-Q_d\sw\right) +
\frac{\sqrt{2}Y_{11}^*Y_{21}}{4m_\om^2G_F} -
\frac{3\sqrt{2}}{32\pi^2G_F}
\frac{(YY^\dagger)_{11}(YY^\dagger)_{21}}{4m_\om^2}, \nonumber \\
g^u_{R} &=& \eta_{R}\left(\frac{1}{4}-Q_u\sw\right) +
\frac{\sqrt{2}F_{11}F^*_{21}}{4m_\chi^2G_F}-
\frac{3\sqrt{2}}{32\pi^2G_F}
\frac{(FF^\dagger)_{11}(FF^\dagger)_{12}}{4m_\chi^2},
\nonumber \\
\Delta g_L &=& \frac{2\sqrt{2}\alpha\pi}{G_Fm_\mu^2}\left(
f_{E0}(-m_\mu^2)+f_{M1}(-m_\mu^2)+f_{M0}(-m_\mu^2)+f_{E1}(-m_\mu^2)\right),
\nonumber \\
\Delta g_R &=& \frac{2\sqrt{2}\alpha\pi}{G_Fm_\mu^2}\left(
f_{E0}(-m_\mu^2)+f_{M1}(-m_\mu^2)-f_{M0}(-m_\mu^2)-f_{E1}(-m_\mu^2)\right)~.
\end{eqnarray}
Here $Q_u$ and $Q_d$ are the charges of the up-- and down--quarks.
As before, we can neglect terms proportional to
$\eta_L$ and $F_3(x_{d_j})$. The summary of constraints from this process
is given in Table
\ref{mu-e-conversion}, with
\begin{eqnarray}
a^L_j &=& (2A-Z)\left[32 \pi^2\delta_{1j}-\yyd\right], ~a^R_j =
2Ze^2 \tilde{h}_j, ~b^L_j = 2Ze^2\tilde{g}_j, \nonumber \\
b^R_j &=& (A+Z)\left[ 32 \pi^2\delta_{1j}-\ffd \right] +
2Ze^2F_4(x_{u_j}) \nonumber \\
&& + ~8\sqrt{2}G_Fm_\chi^2F_5(x_t) \left(\tfrac{3}{4}A-Z\sw \right)
\delta_{3j},
\end{eqnarray}
where $\tilde{g}_{j}=g_{j}|_{q^2=-m_\mu^2}$ and
$\tilde{h}_j=h_j|_{q^2=-m_\mu^2}$. Here we take $|\vec{p}_e|\simeq
E_e \simeq m_\mu$.

\begin{table}[t]
\begin{center}
\begin{tabular}{ccccc}\hline\hline
Element & BR & $Z_{\rm eff}$ & $F_p$ & Constraint \\ \hline \\
$^{48}{\rm Ti}$ & $<4.3 \times 10^{-12}$ &  17.61 & 0.53 &
$\left|\frac{a^L_jY^*_{ij}Y_{2j}}{m_\om^2} + \frac{a^R_j
F_{1j}F_{2j}^*}{m_\chi^2} \right|^2 +
\left|\frac{b^L_jY^*_{ij}Y_{2j}}{m_\om^2} +
\frac{b^R_j F_{1j}F_{2j}^*}{m_\chi^2} \right|^2 < \frac{5.2 \times 10^{-16}}{{\rm GeV}^4}$ \\ \\
$^{208}{\rm Pb}$ & $<4.6 \times 10^{-11}$ & 33.81 & 0.15 &
$\left|\frac{a^L_jY^*_{ij}Y_{2j}}{m_\om^2} + \frac{a^R_j
F_{1j}F_{2j}^*}{m_\chi^2} \right|^2 +
\left|\frac{b^L_jY^*_{ij}Y_{2j}}{m_\om^2} + \frac{b^R_j
F_{1j}F_{2j}^*}{m_\chi^2} \right|^2 < \frac{9.7 \times
10^{-14}}{{\rm GeV}^4}$
\\ \\
\hline \hline
\end{tabular}
\caption{{\footnotesize The summary of $\mu - e$ conversion for Ti
and Pb. The values of $Z_{\rm eff}$ and $F_P$ are taken from
\cite{marciano} whereas $\Gamma_N$ are from \cite{eckhause}.}}
\label{mu-e-conversion}
\end{center}
\end{table}

For $^{48}{\rm Ti}$, we get for LQ mass of 1 TeV, $|Y_{13}^*Y_{23}|
< 4.6 \times 10^{-3}$ and $\left|F_{13}F_{23}^*\right| < 1.9 \times
10^{-4}$ which are slightly stronger than the constraints from
$\mu^- \to e^+e^-e^-$. The tree--level diagrams alone give
$|Y_{11}^*Y_{21}|< 3.3\times 10^{-7}$ and $|F_{11}F_{21}^*| < 3.3
\times 10^{-7}$. A new generation experiment called COMET has been
proposed to reach a better sensitivity of $ 10^{-16}$ \cite{comet}.
There is also discussion of testing $\mu-e$ conversion at a future
Fermilab experiment.  The prospects for these experiments observing
new physics are good within our model.  This is true even if the MEG
experiment \cite{meg} obtains negative results for the $\mu
\rightarrow e \gamma$ decay.\footnote{There is a class of model
where $\mu-e$ conversion is log--enhanced compared to $\mu \to e
\gamma$, see Ref. \cite{sr}.}

We can use the bound from $\mu-e$ conversion and neutrino mass
fitting to predict the minimum branching ratio for $\mu \to 3e$.
Such a correlation is not possible for $\mu \to e \gamma$ because
of the GIM-like cancelation that occurs there, and the fact that
$F_{1j}$ is not constrained by neutrino data.
Note that, $Y_{i3}$ (assuming that only the bottom quark
mediates the process) cannot be smaller than $10^{-4}$ for LQ mass of
300 GeV, or else the induced neutrino mass will be too small. Therefore, the smallest
branching ratio for $\mu \to 3e$ can be predicted which is presented in
Fig. \ref{BR-plot}.  We see that $\mu \rightarrow 3 e$ can be substantial
for a significant part of the parameter space.

\begin{figure}[t]
\centering
    \includegraphics[scale=0.65]{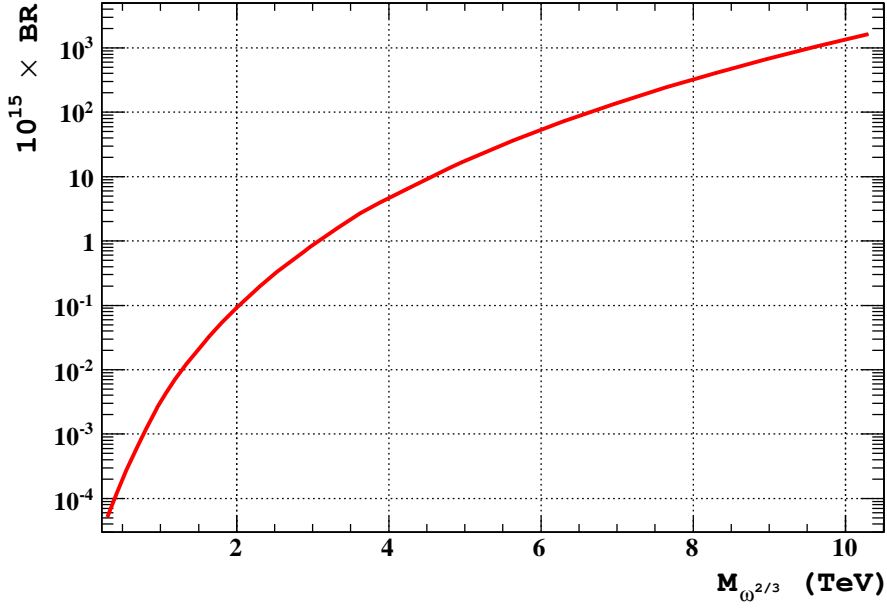}
    \caption{\footnotesize The smallest branching ratio for $\mu \to 3e$ as suggested by neutrino mass fitting
    and $\mu-e$ conversion.}
    \label{BR-plot}
\end{figure}

\subsection{$\bm{\tau^- \to e^-\eta}$ and $\bm{\tau^- \to \mu^- \eta}$ decays}

It turns out that some of the potentially large entries in the neutrino
mass matrix are eliminated from the constraints from the decays
$\tau^- \to e^-\eta$ and $\tau^- \to \mu^- \eta$.  These processes can
be mediated by $\omega^{2/3}$ leptoquark at tree level. Here we
ignore $\eta-\eta'$ mixing and work in the $SU(3)$ flavor symmetric limit.
The branching ratios for these processes are given by
\begin{eqnarray}
{\rm BR}(\tau^- \to e^-\eta) &=& \frac{1}{512
\pi^2}\frac{|Y_{12}Y_{32}|^2}{m_\om^4}\frac{f_\eta^2m_\tau^3}{\Gamma_{\rm total}}, \nonumber \\
{\rm BR}(\tau^- \to \mu^-\eta) &=& \frac{1}{512
\pi^2}\frac{|Y_{22}Y_{32}|^2}{m_\om^4}\frac{f_\eta^2m_\tau^3}{\Gamma_{\rm
total}}.
\end{eqnarray}
We use $f_\eta = 160~{\rm MeV}$ and data from \cite{pdg} to derive
\begin{eqnarray}
|Y_{12}Y_{32}| &<& 1.2 \times 10^{-2} \left( \frac{m_\om}{300~{\rm
GeV}} \right)^2, \nonumber \\
|Y_{22}Y_{32}| &<& 1.0 \times 10^{-2} \left( \frac{m_\om}{300~{\rm
GeV}} \right)^2.
\end{eqnarray}

\subsection{Muon anomalous magnetic moment}

The muon anomalous magnetic moment receives correction from the leptoquark exchange
in our model and is
given by
\begin{eqnarray}
\delta(g-2)_\mu = 3.9\times 10^{-10} \left( \frac{300~{\rm
GeV}}{m_\chi}\right)^2\left( \left|\frac{F_{21}}{1.0}\right|^2 +
\left|\frac{F_{22}}{1.0}\right|^2 -F_4(x_t)
\left|\frac{F_{23}}{1.0}\right|^2\right).
\end{eqnarray}
Again, for the same reason as in the case of $\mu \to e \gamma$,
$Y_{ij}$ couplings are practically have no effects here. By
comparing the new contributions with the $3\sigma$ anomaly in the
experimental value, $\delta(g-2)_\mu^{\rm exp}=(24.6\pm8.0)\times
10^{-10}$ \cite{D-MSSM,anomalous}, we see that this model can reduce
or eliminate the discrepancy for order one values of the couplings
$F_{21}$ and $F_{22}$.

\subsection{Upsilon Decay}

Lepton number violating $\Upsilon$ decay can set limit on the couplings $Y$.
The decays we consider are $\Upsilon \to \mu^+\mu^-$ and $\Upsilon
\to \mu^+\tau^-$. The former is generated by the s-channel photon
exchange and the latter is generated by the t-channel $\om^{2/3}$
exchange. In our model the ratio of these two branching ratios is
given by
\begin{equation}
\frac{{\rm BR}(\Upsilon\to \mu^+\tau^-)}{{\rm BR}(\Upsilon\to
\mu^+\mu^-)} =
\frac{1}{32}\frac{|Y_{23}^*Y_{33}|^2}{16\pi^2\alpha^2}
\left(\frac{m_\Upsilon}{m_\om}\right)^4,
\end{equation}
where $m_\Upsilon$ is the $\Upsilon$ mass. The upper limit of this
ratio is $2.5\times 10^{-4}$ \cite{cleo}. Therefore, we get less
constrained value
\begin{equation}
\left|Y_{23}^*Y_{33}\right| < 8.0 \left( \frac{m_\om}{300~{\rm
GeV}} \right)^2.
\end{equation}

\subsection{$\bm{D_s^- \to \ell^- \nu}$ decay}

Another interesting feature of the leptoquarks is that they can
mediate a lepton number violating $D_s$ decay: $D_s^- \rightarrow
\ell^- \nu$. Currently there is a discrepancy about $2\sigma$ level
between the experimental value and theoretical predictions based on
lattice evaluation of the decay constant $f_{D_s}$.  The LQ exchange
can resolve the discrepancy. This issue has been studied in the
framework of MSSM with broken $R$-parity \cite{D-MSSM,kundu-nandi}
and in extensions of the standard model with leptoquarks
\cite{D-LQ}. It turns out that in this model this process occurs at
tree level mediated by $X_a^{-1/3}$ leptoquarks in addition to the
SM process. Since the SM diagram leads to antineutrino final states,
while the $X_a$ mediating processes result in neutrinos, there is no
interference between the two.\footnote{Analogous lepton number
violating muon decay has been explored as an explanation of the LSND
anomaly in Ref. \cite{babu-pakvasa}.}   We get
\begin{eqnarray}
{\rm BR}(D_s \to \ell_j \nu_i) &=& \frac{m_{D_s}}{8\pi} \tau_{D_s}
f_{D_s}^2 G_F^2|V_{cs}|^2\left[1 +
\frac{|Y_{i2}^*F_{j2}|^2}{128M_1^4G_F^2|V_{cs}|^2}\sin^2 2\theta
\left(1-\frac{M_1^2}{M_2^2}\right)^2\right] \nonumber \\
&&\times \left(1-\frac{m_{\ell_j}^2}{m_{D_s}^2}
\right)^2m_{\ell_j}^2,
\end{eqnarray}
where $m_{D_s}$ and $\tau_{D_s}$ are the $D_s$ mass and lifetime
respectively while $V_{cs}$ is the $(2,2)$ element of CKM matrix.
Comparing results from lattice calculation, $f_{D_s}^{\rm
latt}=(241\pm3)~{\rm MeV}$ \cite{lattice}, and the experimental
result, $f^{\rm exp}_{D_s}=(257.5\pm6.1)~{\rm MeV}$ \cite{rosner},
we see that at present there is a 2$\sigma$ discrepancy.  If this
discrepancy prevails, there is a natural explanation in the present
context. Assuming maximal mixing among the two charge $-1/3$
leptoquarks, we obtain the constraint (corresponding to the central
values of the decay constants $f_{D_s}^{\rm latt}$ and $f_{D_s}^{\rm
exp}$)
\begin{eqnarray}
|Y_{i2}^*F_{j2}| \simeq 2.1 \left(\frac{M_1}{300~ {\rm
GeV}}\right)^2\left(\frac{{\rm 450~GeV}}{\mu}\right).
\end{eqnarray}

There is a similar contribution to the decay $\pi^- \to \mu^- \nu_e$.
From this process we find
\begin{equation}
{\rm BR}(\pi^-\to \mu^-\nu_e) =
\frac{\left|Y_{11}F_{21}^*\right|^2}{128M_1^4G_F^2} \sin^2 2\theta
\left(1-\frac{M_1^2}{M_2^2}\right)^2.
\end{equation}
By using experimental limits \cite{pdg}, one can derive
\begin{equation}
\left|Y_{11}F_{21}^*\right| < 0.62 \left(\frac{M_1}{300~ {\rm
GeV}}\right)^2\left(\frac{{\rm 450~GeV}}{\mu}\right).
\end{equation}

\subsection{New CP violation in $\bm{B_s-\ol{B}_s}$  mixing}

The leptoquarks of the model mediate meson--antimseon mixing via
one--loop box diagrams shown in Fig. \ref{b-bbar}.  There is special interest in this process
for the $B_s$ system, as there are experimental hints for non--zero
CP violation in this system. We find that the present model can
nicely explain the new CP violation that would be needed in
$B_s-\overline{B}_s$ mixing, without generating unacceptable mixing
in the $B_d^0$, $K^0$ and $D^0$ systems.

The new effective Lagrangian for $B_s-\ol{B}_s$ mixing in the model
arising from the box diagrams of Fig. \ref{b-bbar} is given by
\begin{eqnarray}
\mathcal{L}_{\rm eff}^{\rm new} =
-\frac{\left(Y_{i2}Y_{i3}^*\right)^2}{128\pi^2m_\om^2}
\left(\ol{s_R}\gamma^\mu b_R\right)\left(\ol{s_R}\gamma_\mu
b_R\right).
\end{eqnarray}
The $B_s-\overline{B}_s$ transition amplitude is given by
\begin{eqnarray}
\left< B_s\left|-\mathcal{L}_{\rm eff}^{\rm new}\right|\ol{B}_s
\right> =
\frac{\left(Y_{i2}Y_{i3}^*\right)^2}{192\pi^2m_\om^2}~m_{B_s}f_{B_s}^2B_1^{B_s}(\mu)\eta^{B_s}_1(\mu),
\label{Bs-amp}
\end{eqnarray}
where $m_{B_s}$ and $f_{B_s}$ are the $B_s$ meson mass and decay
constant respectively, while $B_1^{B_s}(\mu)$ and
$\eta^{B_s}_1(\mu)$ are the bag parameter and the QCD correction
factor evaluated at the scale $\mu \sim m_b \sim 5$ GeV and their
numerical values are $B_1^{B_s}(5~{\rm GeV})=0.86$ and
$\eta_1^{B_s}(5~{\rm GeV})=0.80$ \cite{becirevic}.

\begin{figure}[t!]
\centering
    \includegraphics[scale=0.8]{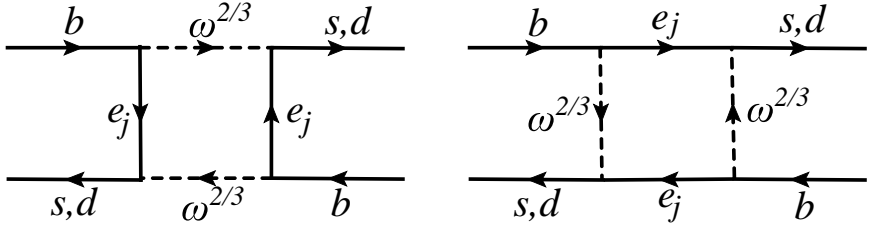}
    \caption{\footnotesize The box diagrams of $B_s-\ol{B}_s$ Mixing.}
    \label{b-bbar}
\end{figure}

Recently the D{\O} collaboration has reported
an excess in the like-sign dimuon asymmetry  \cite{abazov}
defined as
\begin{equation}
A^b_{sl} = \frac{N^{++}-N^{--}}{N^{++}+N^{--}} = -(9.57 \pm 2.51 \pm
1.46) \times 10^{-3}, \label{mu-asym}
\end{equation}
where $N^{++}(N^{--})$ is the numbers of events containing two $b$
hadrons that decay semileptonically into two positive (negative)
muons. Eq. (\ref{mu-asym}) can be written as a linear combination of
two asymmetries \cite{grossman,abazov}
\begin{equation}
A^b_{sl} = (0.506 \pm 0.043) a^d_{sl} + (0.494 \pm 0.043) a^s_{sl},
\end{equation}
where $a^q_{sl}$ ($q\equiv d,s$) is defined as \cite{abazov}
\begin{equation}
a^q_{sl} = \frac{\Gamma(\ol{B}_q \to \mu^+X) - \Gamma(B_{q} \to
\mu^-X)}{\Gamma(\ol{B}_q \to \mu^+X) + \Gamma(B_{q} \to \mu^-X)}.
\label{wc-asym}
\end{equation}
In the SM, $a^d_{sl}=-4.8^{+1.0}_{-1.2} \times 10^{-4}$ and $a^s_{sl}=
(2.1 \pm 0.6)\times 10^{-5}$ \cite{lenz}, so that $(A^b_{sl})^{\rm SM} =
-2.3^{+0.5}_{-0.6} \times 10^{-4}$ which is $3.2\sigma$ away from
the current measurement (see Eq. (\ref{mu-asym})). A likely explanation is that
there is a new source of CP violation in $B_s-\overline{B}_s$ mixing, which
can arise from leptoquarks, as given in Eq. (\ref{Bs-amp}).

The leptoquark contribution will modify the mass difference $\Delta M_s$ in the $B_s$ system
\cite{ligeti}:
\begin{equation}
\Delta M_s = \Delta M_s^{\rm SM} + \Delta M_s^{\rm new} \equiv
\Delta M_s^{\rm SM} \left|1+h_se^{2i\sigma_s}\right|~.
\end{equation}
In our model, the new contribution $h_s$ is given by
\begin{equation}
h_se^{2i\sigma_s} =
\frac{\left(Y_{i2}Y_{i3}^*\right)^2}{384\pi^2m_\om^2M_{12s}^{\rm
SM}}~m_{B_s}f_{B_s}^2B_1^{B_s}(\mu)\eta^{B_s}_1(\mu).
\label{hs}
\end{equation}
The index $i=1-3$ is to be summed in Eq. (\ref{hs}), but in practice
not all three lepton flavors can simultaneously be significant,
owing to lepton flavor violating constraints.  Therefore we focus on the $i = 3$
contributions.

A global fit to the CKM mixing parameters, including the new
D{\O} data on dimuons, finds \cite{ligeti} $\{h_s \sim 0.5,~ \sigma_s \sim 120^o\}$
or \{$h_s \sim 1.8,~\sigma_s \sim 100^o$\} as the best fit values.
The leptoquark contributions will be maximal when
$|Y_{32}| \sim 1$, which is allowed in the model, and when the leptoquark mass
is the smallest ($m_{\omega} \sim 300$ GeV).  In this case $Y_{i2} <  10^{-2}$ for $i=1,2$
in order to satisfy lepton flavor violating constraints.
A fit to the neutrino oscillation data requires all the $Y_{i3}$ ($i=1-3$) to be of the same order
with $|Y_{33}/Y_{23}| \sim 3$ corresponding to $m_\omega=300~{\rm GeV}$.  In this case,
$\mu-e$ conversion in nuclei sets a limit $|Y_{33}| < 0.078$.
Using this upper limit in Eq. (\ref{hs}), and using $|M_{12s}^{\rm SM}| = (9.0 \pm 1.4)~{\rm
ps^{-1}}$ \cite{dobrescu}, we find $h_s \sim 0.42$.  The phase $\sigma_s$ is unconstrained, and
can take the desired value of $120^o$. These are in excellent agreement with the global
fit of Ref. \cite{ligeti}.

We need to point out that the leptoquark model could
affect the absorptive part of the mixing amplitude  \cite{dighe}:\footnote{We thank Amol Dighe for
useful discussion on this point.}
\begin{equation}
\Gamma_{12s} = \Gamma_{12s}^{\rm SM} + \Gamma_{12s}^{\rm new},
\end{equation}
where $\Gamma_{12s}^{\rm new}$ is given by \cite{dighe}
\begin{equation}
\Gamma_{12s}^{\rm new} =
-\frac{\left(Y_{i2}Y_{i3}^*\right)^2}{256\pi
m_\om^4}~m_{B_s}f_{B_s}^2B_1^{B_s}(\mu)m_b^2 F(\ell_i),
\end{equation}
where $F(\ell_i)$ is a kinematic function, as a function of
$m_{\ell_i}^2/m_b^2$. The numerical values of this function are
$F(\ell_i) \simeq 1$ for $\ell_i=e,~\mu$ and $F(\tau) \simeq 0.65$.
By using the same parameter values as for $\Delta M_s$, we see that
the new contribution can modify the SM value of the width difference
by at most 2\%.  There is currently a 2$\sigma$ discrepancy in
$\Delta \Gamma_s$, which should be explained as a statistical
fluctuation within our model.

Similar expressions hold for the mixing of other meson systems:
\begin{eqnarray}
\left< B_d\left|-\mathcal{L}_{\rm eff}^{\rm new}\right|\ol{B}_d
\right> &=&
\frac{\left(Y_{i1}Y_{i3}^*\right)^2}{192\pi^2m_\om^2}~m_{B_d}f_{B_d}^2B_1^{B_d}(\mu)\eta^{B_d}_1(\mu),
\nonumber \\
\left<K\left|-\mathcal{L}_{\rm eff}^{\rm new}\right|\ol{K} \right>
&=& \frac{\left(Y_{i2}Y_{i1}^*\right)^2}{192\pi^2m_\om^2}~m_K
f_K^2B_1^K(\mu)\eta^K_1(\mu),
\nonumber \\
\left< D\left|-\mathcal{L}_{\rm eff}^{\rm new}\right|\ol{D} \right>
&=&
\frac{\left(F_{i1}F_{i2}^*\right)^2}{192\pi^2m_\chi^2}~m_Df_D^2B_1^D(\mu)\eta^D_1(\mu).
\label{meson-amp}
\end{eqnarray}
The $\eta_1$ factors are found by following the procedures given in
Ref. \cite{becirevic} to be $\eta_1^{B_{d}} = 0.80,~\eta_1^K =
0.76,~\eta_1^D = 0.78$, with $B_1^{d}(5~{\rm GeV}) = 0.86,~
B_1^K(2~{\rm GeV}) = 0.66,$~ $B_1^D(2.8~{\rm GeV}) = 0.865$.

The limits on those couplings appearing in $K-\ol{K}$, $D-\ol{D}$,
and $B_d-\ol{B}_d$ mixings can be derived by using $m_K=0.498~{\rm
GeV}$, $f_K = 0.16~{\rm GeV}$, $m_D = 1.9~{\rm GeV}$, $f_D =
0.2~{\rm GeV}$, $m_{B_d}=5.3~{\rm GeV}$, and $f_{B_d}=0.24~{\rm
GeV}$. They are found to be
\begin{eqnarray}
|Y_{i2}Y_{i1}| < 9.6\times 10^{-3} \left(\frac{m_\om}{300~{\rm GeV}}\right), \nonumber \\
|F_{i1}F_{i2}| < 7.3 \times 10^{-3}\left(\frac{m_\chi}{300~{\rm
GeV}}\right), \nonumber \\
|Y_{i2}Y_{i3}| < 1.6\times 10^{-2} \left(\frac{m_\om}{300~{\rm
GeV}}\right).
\end{eqnarray}
These constraints are all consistent with acceptable generation of neutrino masses
and mixings within the model.

\subsection{Neutrinoless double beta decay}

The light neutrino exchange contribution to neutrinoless double beta
decay vanishes in our model, owing to the zero of the (1,1) entry in
$M_\nu$.  However, it can proceed via the vector--scalar exchange
process, which does not need a helicity flip of the neutrino
\cite{vector-scalar}.  The relevant diagram is shown in Fig.
(\ref{double-beta}). The effective Lagrangian of that process after
Fierz rearrangement is
\begin{equation}
\mathcal{L}_{\rm eff} =
\frac{G_F^2}{2}~\epsilon~\ol{u}\ga^\mu(1-\ga_5)d\left[\ol{u}(1+\ga_5)d\ol{e}\gamma_\mu(1-\ga_5)\frac{1}{q\hspace{-6pt}/}e^c
+\frac{1}{4}\ol{u}\sigma^{\alpha\beta}(1+\ga_5)d\ol{e}\gamma_\mu(1-\ga_5)\frac{1}{q\hspace{-6pt}/}\sigma_{\alpha\beta}e^c\right],
\end{equation}
where $q$ is the internal neutrino momentum. The first term is the
scalar-pseudoscalar current and the second one is the tensor
current, and the parameter $\epsilon$ is defined as
\begin{equation}
\epsilon = \frac{Y_{11}^*F_{11}}{2\sqrt{2}M_1^2G_F}\sin 2\theta
\left(1-\frac{M_1^2}{M_2^2}\right),
\end{equation}
where $\theta$ is the leptoquarks mixing angle.
Following P\"{a}s {\it et al.} \cite{pas}, the constraint on
$\left|Y_{11}^*F_{11}\right|$ is calculated by constructing nuclear
matrix element in pn-QRPA model \cite{pnqrpa} and applying the
result of $0\nu\beta\beta$ half life,
$T_{1/2}(0\nu\beta\beta)>1.2\times10^{25}~{\rm yr}$, obtained from
Heidelberg-Moscow experiment \cite{moscow}. A straightforward
calculation gives
\begin{equation}
\left|Y_{11}^*F_{11}\right| < 1.7 \times 10^{-6}\left(\frac{M_1}{1~
{\rm TeV}}\right)^2\left(\frac{{\rm 0.5~TeV}}{\mu}\right).
\label{double-beta-constraint}
\end{equation}
Here, both the scalar-pseudoscalar and tensor currents are taken into
account.  One sees that it is possible to observe neutrinoless double
beta decay in the near future, even with hierarchical neutrino masses.

\begin{figure}[t]
\centering
    \includegraphics[width=4.5cm]{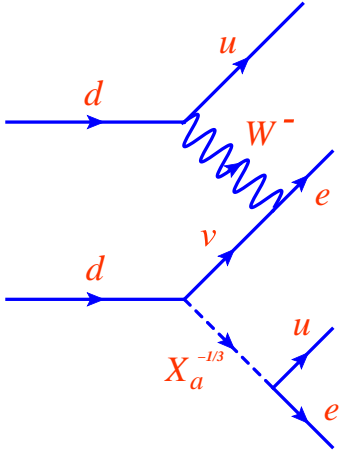}
    \caption{\footnotesize The neutrinoless double beta decay
    diagram generated by this model.}
    \label{double-beta}
\end{figure}

\section{Collider signals}

At hadron colliders, leptoquarks can be produced in pairs via
quark-quark or gluon-gluon fusion and/or in association with lepton
via quark-gluon fusion. The former depends on the QCD coupling
constant and the latter depends on the $|F_{i1}|^2$ and
$|Y_{i1}|^2$. The associated production may be important if the $F$
and $Y$ are as large as electromagnetic coupling $e$ \cite{belyaev}. Since
these couplings are constrained by rare processes such as neutrinoless
double beta decay, it seems
unlikely that single production will be dominant in our model. Therefore, here we
focus on pair production of leptoquarks.

In $\om^{2/3}$ pair production, each will decay to a charged lepton
and a down quark with opposite charges.  The resulting final state is two
leptons and two jets. At the LHC
with $\sqrt{s}=14~{\rm TeV}$ and mass of leptoquark equals to
$500~{\rm GeV}$, the cross section is about $340$ fb  \cite{belyaev},
decreasing to $4.2$ fb for a LQ mass of
1 TeV. The dominant
background for this process is final states from $Z+~{\rm jets}$ and $\ol{t}t$ production.
For high leptoquark mass, only $Z+~{\rm jets}$ is dominant. LHC can
probe this leptoquark mass up to $1.2~{\rm TeV}$ with integrated
luminosity of $300~{\rm fb^{-1}}$.

In $X_a^{-1/3}$ pair production, there are two possible signals. First,
each $X_a^{-1/3}$ decays to up--quark and a lepton, leading to two
lepton and two jets signal. Secondly, one $X_a^{-1/3}$ decays to charged
lepton and up quark and the other one decays to neutrino and a down--type
quark, leading to one lepton plus jets plus missing energy.
The first signature is similar to the $\om^{2/3}$ case. The second
one has dominant background coming from $W+~{\rm jets}$ and
$\ov{t}t$ production. In Ref. \cite{belyaev}, it has been shown that LHC
can probe this leptoquark mass up to $1.2~{\rm TeV}$ with integrated
luminosity of $300~{\rm fb^{-1}}$.

The decays of the $\omega^{2/3}$ leptoquark into $e^+ b,~\mu^+
b,~\tau^+b$ will occur in the following ratios:
\begin{equation}
\Gamma (\omega^{2/3} \rightarrow e^+ b):~ \Gamma (\omega^{2/3}
\rightarrow \mu^+ b):~ \Gamma (\omega^{2/3} \rightarrow \tau^+ b)=
|y|^2:~ |z|^2:~1~. \label{ratio1}
\end{equation}
Measurement of any one of these branching ratios will determine the
CP violating phase $\delta$ via Eq. (\ref{xyz}).  This can of course
be cross checked in long baseline neutrino oscillation experiments,
especially since $\sin^2\theta_{13}$ is large.  If two branching
ratios in the decay of $\omega^{2/3}$ are measured, that will
determine the phase $\delta$ in two different ways, allowing for
another consistency check.

The leptoquarks  $X_a^{-1/3}$, ($a=1,2$) which are linear
combinations of $\chi^{-1/3}$ and $\omega^{-1/3}$ will decay into
charged leptons in the following ratios:
\begin{equation}
\Gamma (X_a^{-1/3} \rightarrow \mu^- t):~ \Gamma (X_a^{-1/3}
\rightarrow \tau^- t) =
 |x|^2~: ~1~.
\end{equation}
Note that this result holds for both $X_a^{-1/3}$, independent of
the $(\omega^{-1/3}-\chi^{-1/3})$ mixing angle $\theta$. Measuring
this branching ratio will determine $|x|$, providing another check
for the model. Since numerically $|x| \gg 1$, we expect that at
least one of the the $X_a^{-1/3}$  will have $\mu^-$ in the final
state dominantly. It should be noted that the decay $X_a^{-1/3}
\rightarrow e^- t$ will be suppressed, owing to constraints from
$\mu \rightarrow e \gamma$. Of course, these $X_a^{-1/3}$ fields
also decay into $\overline{\nu}_i d_j$.  The lepton flavor
composition in this mode would be near to impossible to determine.

\section{Conclusion}

In this paper we have presented a new two--loop neutrino mass generation
model.  This involves TeV scale leptoquarks, which can be directly tested
at the LHC.  The structure of the model is such that the neutrino mixing
angle $\sin^2\theta_{13}$ is predicted to be close to the current experimental
limit of 0.05.  The neutrino oscillation parameters are closely linked to
the decay properties of the leptoquarks.  By measuring the branching ratios
of the LQ bosons, in this model, one is measuring CP violating phase $\delta$
of neutrino oscillations.  These measurements also provide a number of
cross checks by which the model can be falsified.

We have discussed various  rare decay processes mediated by the LQ bosons.
The process $\mu^- \to e^-
\gamma$ may very well be suppressed, owing to a GIM--like mechanism,
while the decay $\mu^- \to e^+e^-e^-$ and $\mu-e$ conversion in nuclei
are within reach of the next generation experiments.  A very interesting feature
of the model is that the present hint of new physics contributions to CP
violation in $B_s-\overline{B}_s$ mixing fits rather well here, without conflicting
other well-measured meson mixings.  Neutrinoless double beta decay can occur
via LQ exchange, even when the neutrino spectrum has a normal hierarchy.
Near future experiments will have a lot to say on the consistency of the model.

\section*{Acknowledgment}

This work is supported in part by US Department of Energy Grant Numbers
DE-FG02-04ER41306 and DE-FG02-ER46140.

\end{document}